\documentclass[a4paper,11pt]{article}
\pdfoutput=1
\usepackage{jheppub} 
\usepackage{graphicx}
\usepackage{amssymb,amsmath,mathrsfs,mathabx}
\usepackage{cancel}
\usepackage{xspace}
\usepackage{subfig}
\usepackage{verbatim}
\newcommand{\Mpl}{M_\mathrm{Pl}}

\newcommand{\GeV}{\mathrm{GeV}}
\newcommand{\MeV}{\mathrm{MeV}}

\newcommand{\keV}{\mathrm{keV}}
\newcommand{\km}{\mathrm{km}}

\newcommand{\beq}{\begin{equation}}
\newcommand{\eeq}{\end{equation}}
\newcommand{\bea}{\begin{eqnarray}}
\newcommand{\eea}{\end{eqnarray}}

\newcommand{\MSbar}{\overline{\mathrm{MS}}}

\newcommand{\Veff}{V_{\mathrm{eff}}}

\newcommand{\meff}{m_{\mathrm{eff}}}

\DeclareMathOperator\heaviside{\theta}

\newcommand{\aeff}{\alpha_{\mathrm{eff}}}

\newcommand{\p}{\varphi}
\newcommand{\pmin}{\varphi_{\mathrm{min}}}
\newcommand{\NS}{\mathrm{NS}}
\newcommand{\WD}{{\mathrm{WD}}}
\newcommand{\Msol}{M_{\odot}}

\newcommand{\triplesys}{\textit{PSR J0337$+$1715}\xspace}
\begin{document}
\preprint{SLAC-PUB-17311}
\title{Consequences of Fine-Tuning for Fifth Force Searches}
\author[a]{Nikita Blinov,}
\author[a]{Sebastian A. R. Ellis,}
\author[b]{Anson Hook}

\affiliation[a]{SLAC National Accelerator Laboratory, 2575 Sand Hill Road, Menlo Park, CA, 94025, USA}
\affiliation[b]{Maryland Center for Fundamental Physics, University of Maryland, College Park, MD 20742}
\date{\today}
\abstract{
  Light bosonic fields mediate long range forces between objects. 
  If these fields have self-interactions, i.e., non-quadratic terms in the potential, 
  the experimental constraints on such forces can be drastically altered due to a 
  screening (chameleon) or enhancement effect.
  We explore how technically natural values for such self-interaction coupling constants modify 
  the existing constraints. We point out that assuming the existence of these natural interactions leads to 
  new constraints, contrary to the usual expectation that screening leads to gaps in coverage.
  We discuss how screening can turn fundamentally equivalence principle (EP)-preserving forces into EP-violating ones.
  This means that when natural screening is present, searches for EP violation can be used to constrain EP-preserving forces. 
  We show how this effect enables the recently discovered stellar triple system \textit{PSR J0337$+$1715} to place a powerful constraint on EP-preserving fifth forces.
  Finally, we demonstrate that technically natural cubic self-interactions modify the vacuum structure of the scalar potential, leading to new constraints from spontaneous and induced 
  vacuum decay.
}
\maketitle

\setlength{\parindent}{4ex}

\section{Introduction\label{sec:intro}}

Modifications of the gravitational inverse-square law provide an important 
test for the existence of large extra dimensions and for the presence 
of new ultra-light bosons that can arise in string theory.  As a result, a range of 
experiments and observations have been carried out to discover 
such forces at a wide range of length scales~\cite{Fischbach:1999bc,Adelberger:2003zx}.
In many cases, experimental constraints exclude fifth forces 
with couplings many orders of magnitude weaker than gravity. 
Taken at face value, these results cast doubt on the potential existence of new light degrees of freedom. 
However, it has been shown that fifth forces generically become screened in the presence of 
  derivative~\cite{Babichev:2009ee} and non-derivative self-interactions~\cite{Khoury:2003aq,Feldman:2006wg,Hinterbichler:2010es}.
In this work we focus on the latter -- the so-called chameleon fields which obtain 
a large mass in a dense object, only allowing a thin shell of the object to source the field.
Experimental limits can be eluded if the screening is severe enough. 
Chameleons, related theories, and their observational implications have been recently reviewed in Refs.~\cite{Burrage:2017qrf,Safronova:2017xyt,Brax:2018,Sakstein:2018}.

At the heart of the chameleon screening effect is the fine-tuning of the vacuum mass of the force
mediator, which enables density-induced effects to drastically alter the shape of the 
scalar potential.\footnote{While we specialize to scalar mediators in this work, we note that screening can also be realized in models with a vector fifth force -- see, e.g., Ref.~\cite{Nelson:2008tn}.}
For example, mass tuning of the QCD axion is known to lead to surprising phenomena in 
finite density environments~\cite{Hook:2017psm}. 
Currently, experimental results in 
searches for long-range forces are presented as constraints on models 
where not only is the mass unnaturally small, but also \emph{all} self-interactions are fine-tuned to zero.
This tuning is not technically natural, since the assumed coupling to matter radiatively 
generates all self-interactions consistent with the symmetries. For large enough couplings to matter, 
these natural values of self-interactions can lead to either screening via the mechanism of Refs.~\cite{Gubser:2004uf,Feldman:2006wg}, or to enhancements, depending on the sign of the self-interaction.\footnote{Interestingly, there exists a new solution to the hierarchy problem that, if applied to fifth force scalars, suppresses all terms in the potential, not just the mass term~\cite{Hook:2018jle} and is not ruled out by data. The phenomenology of fifth forces of this type will be more along the lines of Ref.~\cite{Hook:2017psm} than what we consider here.}

In this paper, we take the self-interactions of the scalar to be of their natural radiatively-generated size. For simplicity, we will consider the scalar to have fundamentally equivalence principle (EP)-preserving interactions with matter. The ``natural'' self-interactions can either close off previously open regions of parameter space or provide even tighter constraints due to either enhancements or tunneling. Surprisingly, we find that if one assumes natural self-interactions, screening does \emph{not} open new gaps in experimental coverage, contrary to the case of $\mathcal{O}(1)$ self-interactions studied in~\cite{Khoury:2003aq,Gubser:2004uf,Feldman:2006wg}. In this work, we explore the landscape of experimental constraints in the presence of these natural couplings, considering only one additional coupling at a time.

We show also how the onset of screening is different for objects with different size and mass. 
This results in effective EP violation, even though the underlying force is fundamentally EP-preserving.
Effective EP violation has been demonstrated previously in the context of chameleon theories in Ref.~\cite{Hui:2009kc}. 
This means that searches for EP-violating forces can be used to constrain EP-preserving forces. 
We focus on planetary and astrophysical systems, because these dense sources can lead to screening even for tiny natural self-interactions.
For example, we consider these effects on searches for deviations in the free fall of the Earth and Moon towards the Sun, and the relative gravitation of the objects in the recently discovered stellar triple system \triplesys~\cite{Ransom:2014xla}. 
These measurements, along with the observation of general-relativistic light 
bending around the Sun~\cite{Bertotti:2003rm}, ultimately place the strongest constraint for scalars with very small masses.

This paper is organized as follows. In Sec.~\ref{sec:screen} we discuss how self interactions lead to enhancements and screening, specializing to simple 
polynomial potentials first studied in Refs.~\cite{Gubser:2004uf,Feldman:2006wg}. 
We consider quantum corrections to the scalar potential and determine the natural size of the self-interactions in Sec.~\ref{sec:natural_self_interactions} 
for generic, dilaton-like and $\mathbb{Z}_N$ scalars. For a generic scalar, the strength of the self-interactions 
grows as its coupling to matter is increased, while the dilaton and $\mathbb{Z}_N$ scalars have naturally small self-interactions.
We investigate the effects of the natural quartic interactions of a generic scalar
in Sec.~\ref{sec:quartic_and_higher}, showing that they can give rise to screening if 
the coupling to matter is large enough. 
The natural size of higher-dimensional (non-renormalizable) terms in the scalar potential is typically too small to give rise to substantial screening.
However, in Sec.~\ref{sec:higherdim} we point out that the discovery of a canonical fifth force can place important constraints on Planck-scale 
quantum gravity contributions to these operators.
The scalar-mediated fifth force in the above cases has the familiar Yukawa-like large-distance behavior.
In contrast, cubic self-interactions, considered in Sec.~\ref{sec:cubic}, are relevant operators which result 
in a different large-distance fall-off. Cubic potential terms can give rise to meta-stable 
vacua, resulting in additional constraints on scalar fifth forces which we also discuss in Sec.~\ref{sec:cubic}.
We summarize these results and conclude in Sec.~\ref{sec:conclusion}.

\section{Screening and Enhancements with Non-Quadratic Potentials\label{sec:screen}}

In this section, we briefly discuss how screening or enhancement occurs in models with non-quadratic potentials.
We consider a scalar field $\p$ with interactions described by 
\beq
\mathscr{L} \supset \frac{1}{2} (\partial \p)^2 - V(\p) + \beta \rho \p ,
\label{eq:generic_fifth_force_lang}
\eeq
where $V$ is the scalar potential, $\rho$ is the matter density and $\beta$ is a coupling constant.
If $\p$ is sufficiently light, it can mediate a new long range force between matter sources $\rho$.\footnote{The coupling to density in Eq.~\ref{eq:generic_fifth_force_lang} can arise in the ultra-violet (UV) from an EP-preserving interaction $\p T$ (where $T$ is the trace of the quantum stress-energy tensor) or via EP-violating, species-dependent interactions. 
For simplicity, we will assume that this interaction is fundamentally EP-preserving.}

The equation of motion (EOM) which determines the profile of $\p$ (and therefore the resulting force on another object) 
due to a spherically symmetric source of constant density is
\beq
\p'' + \frac{2}{r} \p' = V'(\p) - \beta \rho \heaviside(r - R),
\label{eq:eom}
\eeq
where $R$ is the source radius.
For a quadratic potential $V=m^2\p^2/2$ this equation is easily solved to give a Yukawa field profile $\p \sim Q/r$
at distances $m^{-1} \gg r \gg R$, where $Q = \beta M$ is the ``charge'' of the source and $M = 4\pi R^3\rho/3$ is its mass. 
The new $\p$-mediated force between two objects is then simply given by~\cite{Fischbach:1999bc,Adelberger:2003zx}
\beq
F_{5,ij} = \frac{\alpha G M_i M_j}{r^2} \left(1+ m r\right) e^{-m r},
\label{eq:standard_fifth_force}
\eeq
where $\alpha = \beta^2 \Mpl^2/4\pi$, $1/m = \lambda$ is the range of the force, $\Mpl=1.221\times 10^{19}\;\GeV$ is the Planck mass and $G = 1/\Mpl^2$.

The quadratic potential is special because it gives rise to a linear EOM.
Screening or enhancements of the $\p$-mediated force are a generic feature of non-quadratic potentials. 
To study these effects we will consider potentials of the form 
\beq
\label{eq:generic}
V(\p) = \frac{g}{n} \p^n, 
\eeq
where $g$ is a coupling constant and $n$ is a positive integer.

\begin{figure}
\centering
\includegraphics[scale=0.61]{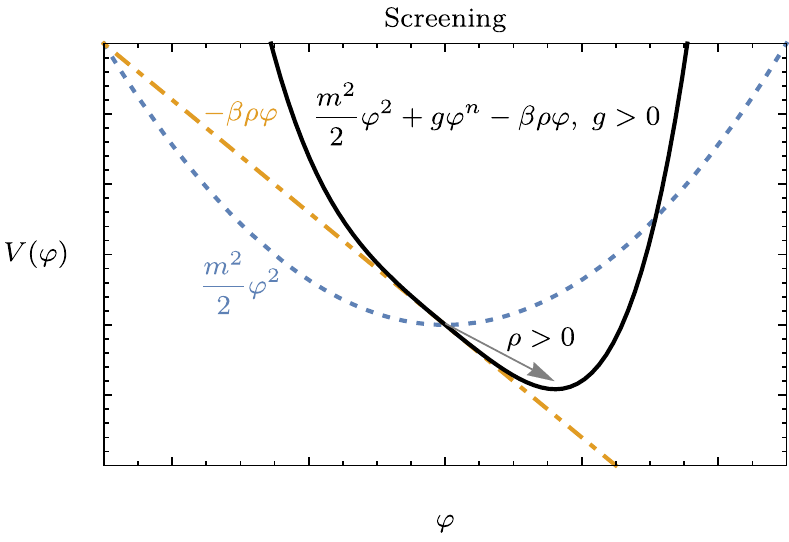}
\includegraphics[scale=0.61]{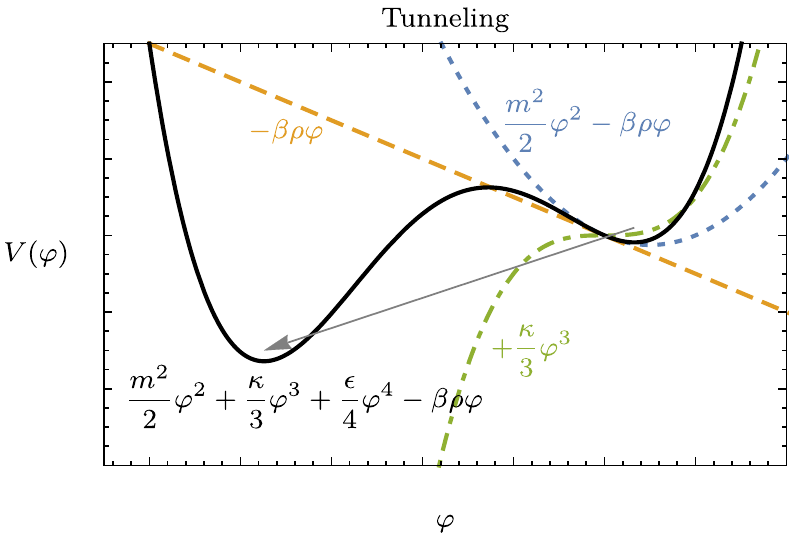}
\includegraphics[scale=0.61]{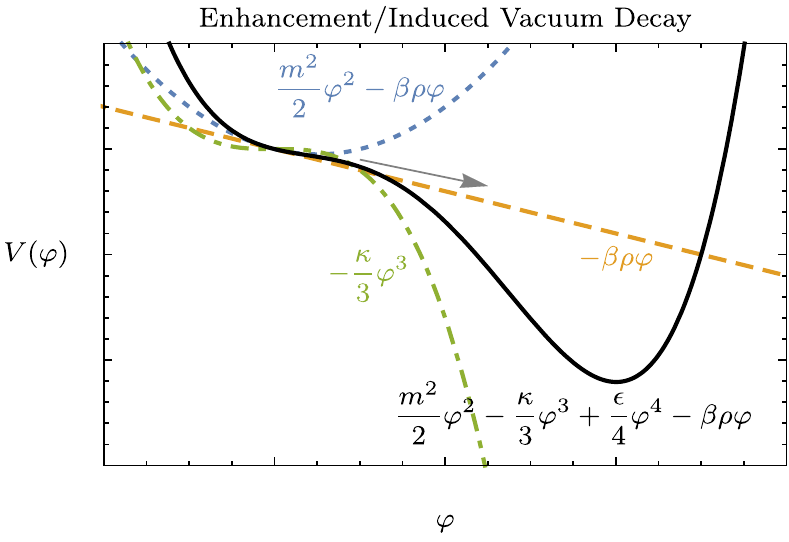}
\caption{Qualitative behavior of the scalar potential $V$ for a fifth force mediated by $\p$ that gives rise to screening (left), tunneling (middle) and enhancement (right).
  In each panel we show the original potential in the presence of only the quadratic term and the coupling to matter (dotted blue), the effect of the coupling to matter (dashed yellow), and the effect of a cubic self-interaction term (dot-dashed green), with the full potential in black. In the left panel, the dotted blue line shows the quadratic term only.
 \textit{Left:} The coupling to matter shifts the minimum, resulting in a steeper potential, reflecting $\meff > m$ inside a dense object. If 
 $\meff^{-1}$ is smaller than the size of the source, its $\p$ charge is screened.
 \textit{Middle:} A cubic term $\kappa \p^3$ in the potential term generates a deep minimum, making our vacuum at $\p =0$ susceptible to tunneling.
 \textit{Right:} If $\beta\kappa < 0$, finite density effects can result in enhancement of the scalar charge if $\p$ attains a value close to the true minimum and 
 returns to $\p \approx 0$ away from the source. If the source is sufficiently dense (or the cubic large enough), $\p$ does not return to $\p = 0$ 
 outside of the source and instead evolves to the true minimum, resulting in classical vacuum decay.
\label{Enhancement.FIG}
}
\end{figure}

\subsection{Screening}

The essence of the chameleon mechanism~\cite{Khoury:2003aq} is the observation that the in-medium mass $\meff$
of $\p$ can be much larger than the vacuum mass determined by $V$. The onset of screening 
occurs when 
\beq
\meff R \gtrsim 1 \ ,
\label{screening.EQ}
\eeq
where $R$ is the size of the source; thus only a 
small fraction of the object is able to source the field.
This regime also corresponds to the field $\p$ in the interior of the source 
  saturating to the minimum value of the effective potential (the last two terms in Eq.~\ref{eq:generic_fifth_force_lang}).

In the screened regime Eq.~\ref{eq:eom} is highly non-linear. While analytic approximations are 
available in various limits~\cite{Mota:2006fz,Mota:2006ed}, in general, the EOM must be solved numerically. 
However, we can roughly estimate the impact of these 
effects as follows. The non-linear terms of the EOM force the field to decrease much faster than $1/r$ until 
they become comparable to the linear terms; this cross-over occurs
outside of the source, but still at distances $r \sim R$.\footnote{This is true for quartic and higher potentials. 
  The cubic is a relevant operator and becomes \emph{more} important at larger distances (until $r\sim m^{-1}$); 
  nevertheless this argument gives reasonable agreement with numerical calculations. We will come 
  back to this later.
} 
After this point, the field profile can be described as 
\beq
\p \sim \frac{Q}{r} \qquad Q = \beta M \gamma,
\label{eq:phi_in_linear_regime}
\eeq
where $\gamma < 1$ is the ratio between the unscreened and screened charges which encodes the effect of non-linearities at large distances. 
We can estimate $\gamma$ by comparing the left- and right-hand sides of the EOM, Eq.~\ref{eq:eom}, at $r\sim R$, i.e. at 
distances where the cross-over between non-linear and linear behavior occurs. In the context of the generic potential shown in Eq.~\ref{eq:generic}, this comparison is between
\bea
\p'' + \frac{2}{r} \p' \sim \frac{Q}{r^3}  \approx g \p^{n-1} \sim g \frac{Q^{n-1}}{r^{n-1}}.
\eea
Solving for $\gamma$ gives the screening factor
\beq
\gamma \sim \left(\frac{g_c}{g}\right)^{1/(n-2)},
\label{GammaScaling.EQ}
\eeq
where 
\beq
g_c = \frac{2 R^{n-4}}{(\beta M)^{n-2}}
\label{eq:critical_screening_coupling}
\eeq
is the critical size of the coupling.
The above is valid when $\gamma \ll 1$, or $g \gg g_c$. 
The suppression factor $\gamma$ is often expressed in terms of the 
internal $\p$ field value ($\pmin$ in this case) and the Newtonian potential 
of the source $\Phi_N = G M/R$~\cite{Khoury:2003aq,Khoury:2003rn}. In this notation, Eqs.~\ref{GammaScaling.EQ} and~\ref{eq:critical_screening_coupling}
become
\beq
\gamma \sim \left(\frac{8\pi}{3}\right)^{1/(n-2)}\left(\frac{\pmin}{\beta\Mpl^2 \Phi_N}\right)^{\frac{n-1}{n-2}}.
\eeq

It is easy to check 
that there are two equivalent interpretations of the screening effect.  The first interpretation is that 
the scalar field inside the source has reached $\varphi\approx \pmin$ minimizing $\Veff$.
Since $\p$ has reached its minimum, $\p$ ceases to change, regardless of the size of the object.
The second interpretation is that there is an 
in-medium mass $\meff^2 = V''(\pmin)$ such that $\meff R \gg 1$, i.e. the ``range'' 
of $\p$ inside the source is smaller than its size. This is illustrated 
in the left panel of Fig.~\ref{Enhancement.FIG} where we show the 
impact of the of the $\p$-matter coupling on the curvature of the potential 
in the presence of self-interactions. If the potential was quadratic,
the vacuum and in-medium $V''$ would be equal.

In the opposite limit, 
$g \ll g_c$, there is no screening and $\gamma \approx 1$. 
Thus, we can define an approximate scaling of $\gamma$ for the entire range of $g$:
\beq
\gamma \sim \min \left[ 1 , \left(\frac{g_c}{g}\right)^{1/(n-2)} \right].
\label{QeffScaling.EQ}
\eeq
This simple argument reproduces the correct scaling for the case of a quartic self-interaction 
obtained in a different way in Ref.~\cite{Feldman:2006wg}, and 
provides good intuition for the numerical results in the following sections. 
From our numerical simulation, we find that while Eq.~\ref{QeffScaling.EQ} gives the correct scaling,
it overestimates the amount of screening by almost an order of magnitude (more sophisticated approximate solutions were derived in Refs.~\cite{Mota:2006fz,Mota:2006ed}).
The force on a test particle far outside of the source is proportional to $\gamma$ and is therefore significantly 
reduced in the screened regime; this leads to a dramatic weakening of experimental 
bounds on scalar fields with this behavior.

Note that $Q$ is independent of $\beta M$ in the screened limit.
In other words, the strength of the $\p$-matter coupling and mass of the object do not actually matter. 
This characteristic 
feature of screened sources occurs because the field falls off fast outside of the 
source due to the large effective mass. It quickly enters the linear regime when the self-interactions 
are barely sub-dominant. As a result, the effective charge of the source is independent of the details of the source itself. 
The initial fast decrease is determined 
by the value of $\p$ inside the source and the corresponding $\meff$, both of 
which depend on $\beta M$. The value of $\p$ at the cross-over between non-linear and linear regimes, however, 
is determined only by the size of self-interaction and not 
by the $\p$ value in the non-linear ``core''.
This can be verified by inserting Eqs.~\ref{GammaScaling.EQ} and~\ref{eq:critical_screening_coupling} into 
Eq.~\ref{eq:phi_in_linear_regime}.
Thus, the large distance behaviour (outside of the core) should 
only depend on $g$~\cite{Mota:2006fz,Mota:2006ed}. This intuition is made 
explicit by the estimate above.

The scaling of the critical self-coupling strength in Eq.~\ref{eq:critical_screening_coupling} 
can be understood as follows. Since $M \sim \rho R^3$, screening is more easily achieved for 
larger objects (for a fixed density $\rho$), as increasing $R$ reduces the gradient energy associated 
with the transition of $\p$ from non-linear to linear evolution and decreases the value
of $\p$ at which this transition occurs.
Similarly, denser objects decrease the $g$ required for screening since 
increasing $\rho$ increases $\meff^2$ for a fixed $R$. 

Finally we briefly comment on how screening does not always lead to weaker constraints. As is well known, as the range of an EP-preserving fifth force gets larger and larger, most of the bounds on it become weaker and weaker. This is because in this limit, it is difficult to differentiate between the presence of a fifth force and a shift in the mass of the object\footnote{The bending of light around a massive object is one of the few experiments that can differentiate between the two.}. This changes when screening occurs because two objects with the same mass but different radii can have different screening parameters $\gamma$, leading to effective EP violation~\cite{Hui:2009kc}. Since the screening parameters $\gamma$ are independent of the vacuum mass 
of the fifth-force mediator (as long as it is smaller than $R^{-1}$), this effective EP violation persists for arbitrarily long-ranged forces.
We will show this explicitly in Sec.~\ref{sec:quartic_and_higher}.

\subsection{Enhancements and Vacuum Decay}
Screening takes place when $\pmin\ll \beta M/R$, 
i.e. when $\p$ at the true minimum of the effective potential is smaller than the surface value of $\p$ in the absence of 
self-interactions. This is equivalent to the condition in Eq.~\ref{screening.EQ}. When the opposite is true, $|\pmin|\gg \beta M/R$, two new effects can occur depending 
on the shape of the potential. First, the vacuum near $\p \approx 0$ can be metastable and therefore 
susceptible to quantum tunneling. This is shown in the middle panel of Fig.~\ref{Enhancement.FIG}.
The second possibility is that the scalar charge of an object can be enhanced.
Enhancements occur when finite density effects encourage the growth of $\p$ (e.g. a negative mass squared at small $\p$) rather than discourage the growth of $\p$ (e.g. a positive mass squared at small $\p$).  
When $\pmin$ is large enough, enhancement can induce classical vacuum decay. We illustrate 
  the potential that gives rise to enhancement in the right panel of Fig.~\ref{Enhancement.FIG}.
For the remainder of this section we focus on enhancement and induced vacuum decay since 
these will provide the strongest constraints in Sec.~\ref{sec:cubic}.

As in the case of screening, $\meff$ at the true in-medium minimum is larger than the vacuum mass $m$. 
This would lead one to expect that for a large object the force should be weaker. 
However, unlike the case of screening, $\pmin$ can be parametrically larger than $\beta M/R$. 
As a result, the field profile inside the source starts at a much higher value, so that $\p(r)$ outside 
the source is much larger than it would be if the self-interaction terms were absent. 
Therefore, when discussing enhancements, it is more useful to think about 
the screening/enhancement factor $\gamma$ and $\p$ at the potential minimum rather than the effective mass. 

The concrete case we will consider is a potential with a coupling of $\p$ to matter, a cubic self-interaction term, and a small quartic self-interaction term to stabilize the potential
 \beq
 V(\p) = \frac{m^2}{2}\p^2 + \frac{\kappa}{3}\p^3+\frac{\epsilon}{4}\p^4 - \beta \rho \p \ ,
 \eeq 
shown in Fig. \ref{Enhancement.FIG}.
In the absence of the quartic, we would find that if
 \bea
\kappa \beta < -\frac{m^4}{4 \rho} \ ,
\eea
the potential would have a runaway direction $\p \to \infty$ and no minimum. 
Unlike the case where there is only a linear term in the EOM (giving rise to the standard Yukawa potential), a cubic term causes a runaway that is faster than exponential and so $\p$ reaches infinity in finite distance.  Thus we include a small quartic term to stabilize the potential, so that $\p$ would instead evolve to the global minimum. With a stabilized potential as above, there are two qualitatively different scenarios, depending on the relative sign of $\beta\kappa$. 

If $\beta\kappa > 0$ screening occurs. Since the $\p$-matter coupling pushes $\p$ to positive values, the self-interaction increases the 
the effective mass at the in-medium minimum. However,  the minimum near $\p \approx 0$ is local both in vacuum and at finite density. This vacuum 
is therefore metastable due to possible tunneling to the global minimum at $\p < 0$; this is shown in the middle panel of Fig.~\ref{Enhancement.FIG}.

To see the simplest example of an enhancement, we consider $\beta\kappa < 0$ as in the right side of Fig. \ref{Enhancement.FIG}. 
In this case $\p$ will evolve classically towards its global minimum, and because a cubic term gives faster than exponential growth it can very quickly reach the minimum. 
Because the position of the global minimum is sensitive to the quartic coupling, the magnitude of the enhancement factor is large, but UV-sensitive. 
As a result, when the force is enhanced due to $\p$ reaching its global minimum, it is difficult to determine whether or not this possibility is excluded in a given observation. 

The enhancements discussed above are similar to the enhancements due to spontaneous scalarization discussed in Ref.~\cite{Damour:1993hw}.  In both cases, the scalar is pushed away from the origin due to finite density effects.  Spontaneous scalarization utilizes a finite density negative mass term while we utilize a cubic term.  Unsurprisingly, a higher power of $\p$ results in a larger effect that occurs for all values of the coupling.  This is in contrast to spontaneous scalarization which turns on only for objects whose densities/radii exceed a critical density/radius.

The assumption that there is a large cubic self-interaction means that the vacuum is in a metastable state regardless of the sign of $\beta\kappa$.
For $\beta\kappa > 0$, this leads to quantum tunneling, as described above.
If, on the other hand, $\beta\kappa < 0$ and an object sources a large field value, it has the potential to create a bubble of true vacuum that will devour the
entire rest of the universe. This and other tunneling constraints can be used to place extremely tight bounds on the scenario of enhancement.
We consider these effects in the context of the cubic self-interaction in Sec.~\ref{sec:cubic}.

\section{Naturalness and Self-Interactions\label{sec:natural_self_interactions}}

Quantum corrections in \emph{perturbative} quantum field theory are small and, therefore, usually provide small 
shifts in relationships between various observables. The situation is different 
when one or more parameters are tuned to be small without the protection of a symmetry. In this situation,  
these small quantum shifts can qualitatively change the behavior of a given theory. A famous example of this 
is spontaneous symmetry breaking in the Coleman-Weinberg model with a classically massless scalar~\cite{Coleman:1973jx}.
In the chameleon theories introduced in the previous sections there are two sources of 
quantum corrections: the $\p$ self-interaction and the $\p$ coupling to Standard Model (SM) fields.
The effect of the former was recently studied in Ref.~\cite{Upadhye:2012vh}. Here we consider 
the latter possibility.\footnote{Recently, Ref.~\cite{Burrage:2016xzz} has investigated the impact of 
quantum corrections due to non-SM matter in the symmetron models.} We first consider the impact of quantum corrections for a generic 
scalar field with no protective symmetries, and show that for a sufficiently large 
coupling to matter these corrections lead to screening. We then contrast these results 
with a dilaton field whose interactions with matter are governed by conformal invariance and $\mathbb{Z}_N$ scalars whose interactions are governed by a $\mathbb{Z}_N$ symmetry.

\subsection{Generic Scalars}
The coupling to matter in Eq.~\ref{eq:generic_fifth_force_lang} induces quantum corrections to $V$ which 
generates all $\p$ self-interactions consistent with symmetries. 
We estimate the size of these corrections as follows.
First, it is useful to write down the $\varphi$-matter coupling as the Yukawa interaction
\beq
\mathscr{L}\supset - (m_f - y\varphi) \bar f f,
\label{eq:yukawa_interaction}
\eeq
where $m_f$ is the fermion mass and  $y = m_f \beta$.
At low energies the relevant fermions are $f = p$, $n$ and $e$, with the coupling to bulk matter in Eq.~\ref{eq:generic_fifth_force_lang} 
dominated by the nucleon couplings. Given a tree-level potential of the form 
\beq
V(\p) = \frac{1}{2} m^2 \p^2 + \frac{1}{3}\kappa \p^3 + \frac{1}{4}\epsilon\p^4,
\label{eq:tree_level_potential}
\eeq
the coupling to matter in Eq.~\ref{eq:yukawa_interaction} gives rise to renormalization group (RG) evolution 
of the $\p$ self-interactions. In the leading-log approximation we find 
\begin{align}
\kappa(\mu) & \approx \kappa (\mu_0) + \frac{3y^3 }{2\pi^2} m_f\ln\frac{\mu}{\mu_0} \\
\epsilon(\mu) & \approx \epsilon(\mu_0) - \frac{y^4}{2\pi^2} \ln\frac{\mu}{\mu_0}. 
\label{eq:potentian_parameter_rg}
\end{align}
Thus, the running of cubic and quartic interactions implies an \emph{approximate} minimum natural bound 
on the size of the self-interactions
\begin{align}
|\kappa| & \gtrsim \frac{3y^3 }{2\pi^2} m_f \\
|\epsilon| & \gtrsim \frac{y^4}{2\pi^2},
\end{align}
where we take the logarithm to be of order one.
For simplicity, we will take the sign of $\epsilon$ to be positive to ensure the potential is bounded from below at the renormalizable level $\mathcal{O}(\p^4)$.\footnote{The potential can have a stable minimum with a negative quartic if higher order terms become important as is the case for a cosine potential.}
The quartic and cubic potential parameters can be made much smaller, but this 
requires the tuning of input parameters against loop corrections, 
\emph{in addition} to the tuning required to keep $\p$ light.
The latter tuning is the usual naturalness problem associated 
with fundamental scalar fields. 

Aside from the renormalizable couplings in Eq.~\ref{eq:tree_level_potential}, 
the $\p$-matter interaction generates all self-interactions $g \p^n$.
These corrections are readily computed at one loop using the Coleman-Weinberg (CW) approach (see Refs.~\cite{Coleman:1973jx,Quiros:1999jp}).
The Yukawa interaction in Eq.~\ref{eq:yukawa_interaction} leads to the $\p$ potential
\beq
V_{\mathrm{CW}} = -\frac{1}{16\pi^2} m_f(\varphi)^4\left(\ln \frac{m_f(\varphi)^2}{\mu^2} - \frac{3}{2}\right),
\label{eq:cw_potential}
\eeq
where $m_f(\p) = m_f - y \p$ is the field-dependent mass of $f$ and $\mu$ is the $\MSbar$ renormalization scale.
Expanding $V_{\mathrm{CW}}$ around $\p=0$, we find the characteristic size of one-loop corrections to be 
\beq
g \sim \frac{y^n m_f^{4-n}}{16\pi^2} = \frac{\beta^n m_f^4}{16\pi^2}.
\label{eq:natural_self_interaction}
\eeq
We note that this is a conservative estimate: a generation-universal, EP-preserving coupling to 
$T_\mu^\mu$ implies interactions with a large number of heavier fermions and gauge bosons which will also 
contribute to RG evolution of self-interactions. The large multiplicity of 
particles running in the loop tends to increase the magnitude of these corrections, barring 
precise cancellations between fermionic and bosonic contributions.
This definition of a ``natural'' self-interaction differs from 
those used previously in, e.g., Refs.~\cite{Gubser:2004uf,Mota:2006fz,Mota:2006ed}, where 
quartics of $\mathcal{O}(1)$ were considered natural. Our minimal-sized 
coupling, Eq.~\ref{eq:natural_self_interaction}, is more in line
with 't Hooft's notion of naturalness~\cite{tHooft:1979rat}, since 
in the limit $y\rightarrow 0$, the vanishing self-interactions 
result in an enhanced symmetry of the theory -- conserved $\p$ number.

The size of the natural self-interaction in Eq.~\ref{eq:natural_self_interaction} grows with the $\p$-matter coupling, so it is clear 
that it will become important at some point. Below we show that these self-interactions  
lead to screening for experimentally interesting values of $y$, or equivalently, of $\alpha$. 

\paragraph{Threshold for Screening}

Given the natural size of the self-interactions described above we can estimate 
when screening becomes important for a given self-interaction term and source object. 
The general screening condition given in Eq. \ref{screening.EQ} can be re-written 
by solving for $\meff$ induced by $V = g \p^n/n$ with $g$ given by Eq.~\ref{eq:natural_self_interaction}. 
This, in turn, allows one to define a critical coupling strength $\alpha^{(n)}_{c,i}$ at which $\meff = 1/R_i$, 
signifying the onset of screening for a given object $i$ in the presence of an $n$-th order self-interaction term. 

The effective mass in an object of density $\rho_i$ is
\beq
\meff^2 \sim \beta^{(2n-2)} m_f^4 \left(\rho_i R_i^2 \right)^{(n-2)}
\eeq
in the limit where the vacuum mass is small.
The screening condition $\meff R_i > 1$ then translates into a bound of $\alpha>\alpha^{(n)}_{c,i}$ where 
\beq
\alpha^{(n)}_{c,i} \sim \frac{\Mpl^2}{R_i^2 \left(m_f^4 \  \rho_i^{n-2} \right)^{1/(n-1)}} \,.
\label{eq:critical_alpha}
\eeq
The critical values of $\alpha$ for various sources considered below are listed in Table \ref{alphaCrit.TAB}. We find that these estimates, together with the approximate scaling of the screening parameter $\gamma$ as discussed around Eq.~\ref{QeffScaling.EQ}, are good approximations of the final numerical results. The critical values of $\alpha$ for the sources considered here correspond the $\p$-matter Yukawa coupling strength $y \ll 1$ (defined in Eq.~\ref{eq:yukawa_interaction}), so the quantum corrections remain under control.

It is important to note that objects with different sizes and masses have different values $\alpha_c$. 
As a result, a force which is fundamentally EP-preserving will appear to see unequal charges from the two objects, resulting in apparent EP violation. For example, due to the different sizes of the Moon and the LAGEOS satellite, there is a large range of $\alpha$ (see Tab.~\ref{alphaCrit.TAB}) for which they experience different accelerations towards the Earth. These differential accelerations are tightly constrained, giving rise to important bounds that we consider in the following sections.

From Eq. \ref{eq:critical_screening_coupling}, we see that the critical coupling $g_c$ above which there is screening depends on the geometry of the object, so that regardless of how $g$ is generated, there will be differences in the screening of different objects. Thus, the argument of effective EP violation arising for screened EP-preserving forces does not depend on our premise of natural-sized couplings.

In the following sections, we consider the impact of screening at $\alpha > \alpha_c$ due the natural-sized self-interactions in Eq.~\ref{eq:natural_self_interaction}, on existing constraints from experimental searches for fifth 
forces. Before this, however, we discuss two exceptions to the preceding analysis. 

\renewcommand{\arraystretch}{1.3}
\begin{table}[t]
\centering
\begin{tabular}{|c | c | c | c|}
  \hline
\textbf{Object} & $\alpha^{(3)}_c$ & $\alpha^{(4)}_c$ & $\alpha^{(5)}_c$ \\
\hline
\hline 
Earth ($\Earth$) & $10^{2}$ & $10^{4.1}$ & $10^{5}$ \\
\hline
Moon ($\Moon$) & $10^{3.2}$ & $10^{5.4}$ & $10^{6.3}$ \\
\hline
Mercury ($\Mercury$) & $10^{2.8}$ & $10^{5}$ & $10^{5.9}$ \\
\hline
Mars ($\Mars$) & $10^{2.6}$ & $10^{4.8}$ & $10^{5.7}$ \\
\hline
LAGEOS ($L$) & $10^{17}$ & $10^{19}$ & $10^{20}$ \\
\hline
Sun ($\Sun$) & $10^{-1.8}$ & $10^{0.44}$ & $10^{1.4}$ \\
\hline
Pulsar ($P$) & $10^{0.48-0.55}$ & $10^{0.34-0.35}$ & $10^{0.09-0.13}$ \\
\hline
Inner Dwarf ($D_i$) & $10^{-0.53}$ & $10^{0.92}$ & $10^{1.5}$ \\
\hline
Outer Dwarf ($D_o$) & $10^{-0.64}$ & $10^{0.70}$ & $10^{1.2}$ \\
\hline
\end{tabular}
\caption{Critical values $\alpha_c$ of the force strength relative to gravity for natural-sized scalar cubic, quartic and quintic self-interactions.
  For $\alpha > \alpha_c$ the scalar fifth force becomes screened. Due to uncertainty in the neutron star equation of state, we provide a range of $\alpha_c$ for the pulsar in \triplesys, where we have used the upper and lower bounds on the radius of a neutron star of mass $1.4 M_\Sun$, $9.9 \lesssim R \lesssim 13.6$ km, from \cite{Annala:2017llu}.
}
\label{alphaCrit.TAB}
\end{table}

\subsection{Dilatons}
In the generic scalar model considered above, vanishing of interactions results in an enhanced symmetry ($\p$ 
number conservation); the mass, on the other hand, is not protected by any symmetry and therefore must be tuned to keep 
$\p$ light.   An interesting counterexample to the previous estimates of natural sizes of couplings occurs for the dilaton.

Technically natural scalar 
fields arise as Nambu-Goldstone bosons (NGBs) of spontaneously broken global symmetries. A small mass 
for these scalars is generated by an explicit breaking of the symmetry.
The vanishing of the mass restores the full (non-linearly realized) symmetry. This symmetry typically 
forbids non-derivative self-interactions. If these are generated, they must be proportional to the 
explicit breaking parameter, or equivalently, the field mass. Thus, a small mass generally implies 
small self-interactions. 

This idea is explicitly realized in the context of spontaneously-broken 
conformal invariance, where $\p$ is the dilaton, the NGB of dilatation symmetry.
Unlike NGBs associated with internal symmetries, the conformal symmetry 
allows for $\p^4$ self-interactions. It is therefore interesting to ask whether 
these can be large enough to enable the chameleon mechanism.
The way in which the dilaton circumvents the previous arguments for large self-interaction terms is 
that a large quartic is generated but is dynamically relaxed to close to zero. Because the dilaton is the NGB of dilatation symmetry, 
it appears in the renormalization logarithms as well as outside of 
the logs. Thus quantum effects serve only to generate a sizeable quartic coupling. In the absence of explicit 
breaking of dilatation symmetry, the minimum is either at $\p = 0$, where dilatation symmetry is not 
spontaneously broken, or $\p \rightarrow \infty$, where dilatation symmetry was never a good symmetry to begin with.
In order to obtain a finite value for the dilatation symmetry breaking scale $f$, an explicit breaking 
must also be introduced. This breaking then generates a non-trivial potential for the dilaton.

The dilaton potential, and therefore its self-interactions, can be constructed 
in the limit of small explicit breaking of the conformal symmetry~\cite{Rattazzi:2000hs,Goldberger:2008zz,Chacko:2012sy}.
Fixing the scale of conformal symmetry breaking $f$ and dilaton mass $m \ll f$ 
one finds 
\beq
V \approx \frac{1}{2}m^2 \p^2 + \frac{a m^2}{f} \p^3 + \frac{b m^2}{f^2}\p^4 + \dots, 
\label{eq:dilaton_potential}
\eeq
where $a$ and $b$ are $\mathcal{O}(1)$ coefficients that can be computed as an expansion in the explicit 
breaking parameter~\cite{Goldberger:2008zz}. 
For example, if the explicit breaking is due to an operator of the form 
$\p^4 (\p/f)^{\Delta - 4}$, then in the limit where $|\Delta-4| \ll 1$, we have 
$a = 5/6$ and $b=11/24$. The cubic and quartic around the minimum are small even though the original quartic was very large.
Thus the $\p$ self-interactions are naturally suppressed 
by $m/f\ll 1$, so, unlike the generic scalar field considered before, a parametrically small mass implies 
parametrically small self-interactions.

If the SM is part of the conformal sector from which the dilaton originates, the couplings of $\p$ 
with the SM are determined by the Noether theorem to be 
\beq
\mathscr{L} \supset \frac{\p}{f} T^\mu_\mu \rightarrow \frac{\p}{f}\rho  
\label{eq:dilaton_matter_coupling}
\eeq
where $T$ is the SM stress-energy tensor (including quantum corrections). Thus $\p$ interactions are 
naturally EP-preserving, and, in the notation of Eq.~\ref{eq:generic_fifth_force_lang}, $\beta=1/f$.
For the small $\p$ masses that we consider in this paper, the cubic and quartic interactions of the dilaton will be
small enough that they do not lead to chameleon-like effects.

\subsection{$\mathbb{Z}_N$ scalars}

Like dilatons, $\mathbb{Z}_N$ scalars are another example of how a symmetry can be used
to suppress both the mass term and higher order self interactions~\cite{Hook:2018jle}. The $\mathbb{Z}_N$ symmetry is non-linearly realized on the scalar as a shift symmetry, and as an exchange symmetry on the $N$ copies of particles with which it interacts. For our purposes, $N$ can be thought of as enumerating copies of the Standard Model. 

If $\p$ has a spurion $\varepsilon$ which breaks the arbitrary shift symmetry $\p \to \p + \theta$ down to $\p \to \p + 2\pi f$, then $\p$ only appears in the potential as
\beq
\mathscr{L} \supset \varepsilon \sin \left( \frac{\p}{f} + \vartheta \right) \ .
\eeq
This theory also has a $\mathbb{Z}_N$ shift symmetry, under which $\p \rightarrow \p + \frac{2 \pi f}{N}$. However, the spurion is not invariant under this symmetry. Utilizing these ingredients, 
the entire potential for $\p$ can be suppressed. $N$ is in principle any integer greater than $2$.

The potential and interactions of $\mathbb{Z}_N$ scalars with matter fermions $\psi_k$ scale as
\bea
\mathscr{L} \sim \sum_k^N \varepsilon \sin \left ( \frac{\p}{f} + \frac{2 \pi k}{N} \right ) \overline \psi_k \psi_k  + \left(\frac{\varepsilon}{m_\psi}\right)^N  m_\psi^4 \cos \frac{N \p}{f} \ .
\eea
The fifth force scalar $\p$ therefore interacts with the $N$ different mirror worlds, each one with its own fermion $\psi_k$, with slightly different couplings.  Due to various trigonometric identities, or more generally the exponential convergence of Riemann sums of periodic functions, the potential for $\p$ is exponentially suppressed by $\varepsilon^N$. 
In practice $N=3$ is good enough to achieve significant suppression of the potential.

What is of relevance for fifth forces is the Yukawa interaction and the mass potential. Ignoring
the mirror world particles and taking the leading order interactions of $\p$, we get the potential 
\bea
\mathscr{L} = \beta \rho \p + \frac{m^2 f^2}{N^2} \cos \frac{N \p}{f}
\eea
Unlike the case of the dilaton, $\beta$ is not entirely fixed by $f$ ($\beta \sim \varepsilon/( m_\psi f )$).  There are three parameters that characterize $\mathbb{Z}_N$ scalars ($f$, $\varepsilon$ and $N$) as opposed to the two parameters that characterize dilatons ($f$ and $m$).  As a result, $\mathbb{Z}_N$ scalars have more flexibility.  $\mathbb{Z}_N$ scalars can interpolate between having self interactions too small to be observed to large self interactions that cause screening.

\section{Quartic Self-Interactions\label{sec:quartic_and_higher}}
\begin{figure}
  \centering
  \includegraphics[width=7.5cm]{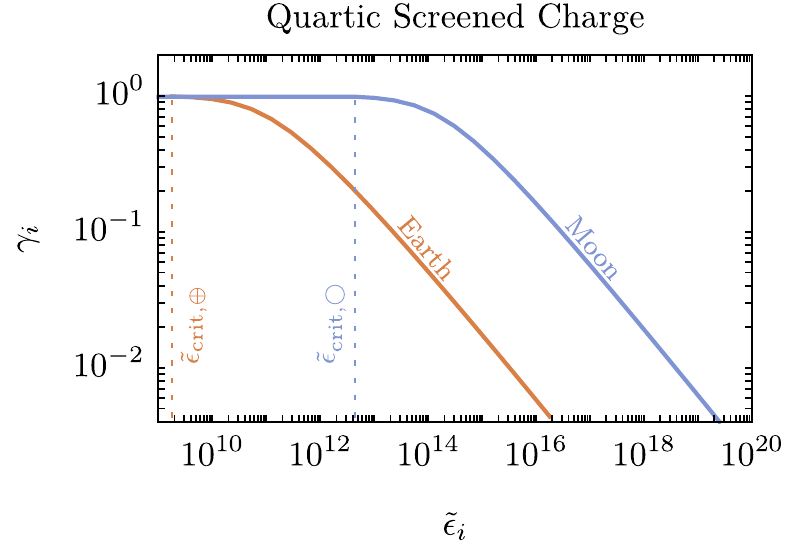}
  \includegraphics[width=7.5cm]{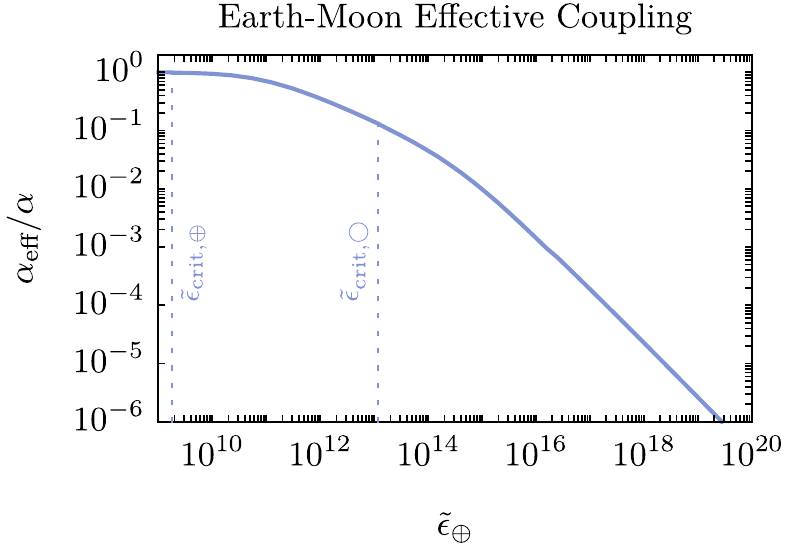}
  \caption{
    Effective charges of the Earth and the Moon (left panel) and the resulting effective 
    coupling (right panel) as a function of the $\p$ quartic self-interaction $\epsilon$ 
    (in units of $(\beta \rho_i/m_0^3)^2$, with $\rho_i$ the density of the source and $m_0^{-1}$ set to 
    the Earth-Moon distance). Experimental bounds on the fifth force coupling are 
    significantly weakened when screening becomes effective inside the Earth and the Moon. 
    These ``thresholds'' are indicated by vertical dashed lines in each panel. They 
    correspond to the effective mass of $\p$ becoming comparable to the source size. 
    \label{fig:quartic_screened_charges}
  }
\end{figure}

As discussed in Sec.~\ref{sec:screen}, in the presence of $g \p^n$, $n\geq 4$ interactions 
in the $\p$ potential, the field far outside of a source is Yukawa-like, with 
an amplitude that is described by an effective charge determined by the thin shell mechanism.
The total force between two such (well-separated) objects $i$ and $j$ is then a simple generalization of the 
usual fifth force parametrization~\cite{Fischbach:1999bc}
\beq
F_{ij} = \frac{G M_i M_j}{r^2} \left[1 + \aeff \left(1+ m r \right) e^{-m r} \right] \ ,
\label{eq:effective_fifth_force}
\eeq
where the effective coupling strength $\aeff$ is
\beq
\aeff= \alpha \gamma_{i}(g) \gamma_{j}(g).
\label{eq:aeff_def}
\eeq
Below we demonstrate how a technically natural value of the quartic coupling 
strength $g=\epsilon$ (see Eq.~\ref{eq:tree_level_potential})
modifies the existing constraints on new long-range forces. 
In Fig.~\ref{fig:quartic_screened_charges} we show how variations of $\epsilon$ 
modify the screening parameters $\gamma_i$ for Earth and Moon (left panel) and 
resulting effective coupling strength $\aeff$ (right panel).
The latter is the quantity (rather than $\alpha$) that is directly 
constrained by, e.g., anomalous lunar precession, as we describe below.

\begin{figure}
  \centering
  \includegraphics[]{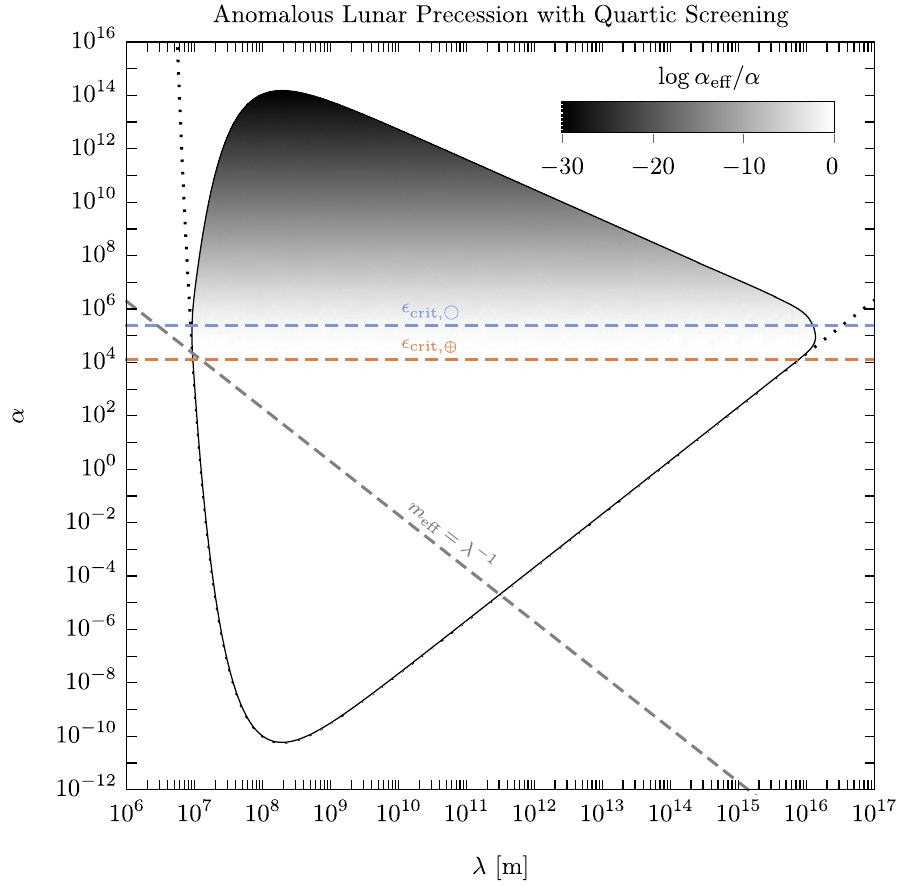}
  \caption{
    Anomalous lunar precession bound on $\p$-mediated fifth force in the plane of coupling strength $\alpha$ and range $\lambda$ 
    including the effects of a natural quartic self-interaction $\epsilon\; \propto\;\alpha^2$.
    The area inside of the black contour is excluded by the existing bound on the anomalous precession of the 
    Moon. The lower edge of the exclusion matches the un-screened case (dotted black line). As one goes 
    up in $\alpha$, the natural quartic becomes large enough to begin to screen the 
    effective charges of the Earth and then the Moon (lower and upper horizontal dashed lines, respectively).
    The reduction of the effective coupling $\aeff$ due to screening is shown by the 
    density plot, with darker colors corresponding to largest screening/smallest $\aeff$.
    Above the gray diagonal dashed line the contribution of the quartic term in the 
    potential to the $\p$ mass in Earth is larger than the vacuum mass $\lambda^{-1}$.
    \label{fig:quartic_screened_llr}
  }
\end{figure}

As argued in the previous section, in the absence of symmetries the \emph{minimal} 
natural size for a quartic self-interaction is parametrically
\beq
\epsilon \sim \frac{\alpha^2 m_n^4}{\Mpl^4}.
\eeq
Thus the self-interaction becomes more important as the scalar field's coupling to matter, $\alpha$, 
is increased. We illustrate the impact of this natural quartic on a representative experimental 
constraint on new fifth forces in Fig.~\ref{fig:quartic_screened_llr}, where we show 
how the anomalous lunar precession bound is modified. This measurement is described in more detail below.
The dotted black line shows the constraint in the absence of screening. Above the dashed line labelled $\meff>\lambda^{-1}$, the in-medium mass $\meff$ is greater than the bare mass given by $\lambda^{-1}$, and an object of size $R = \lambda$ will therefore be screened. For example, one can see that $\epsilon_{\text{crit},\Moon}$ intersects this line roughly at $\lambda=R_\Moon$. We see that above the horizontal lines denoting $\alpha_{c,\Earth}$ and $\alpha_{c,\Moon}$, $\alpha_{\text{eff}}/\alpha$ shrinks due to the screening parameters $\gamma_\Earth$ and $\gamma_\Moon$ decreasing as a function of increasing $\alpha$, until eventually $\aeff(\lambda)$ satisfies the experimental constraint at large values of $\alpha$. Thus the previously unbounded-above experimental constraint now only excludes a finite region of the $\alpha-\lambda$ parameter space.

\begin{figure}
  \centering
  \includegraphics[width=10cm]{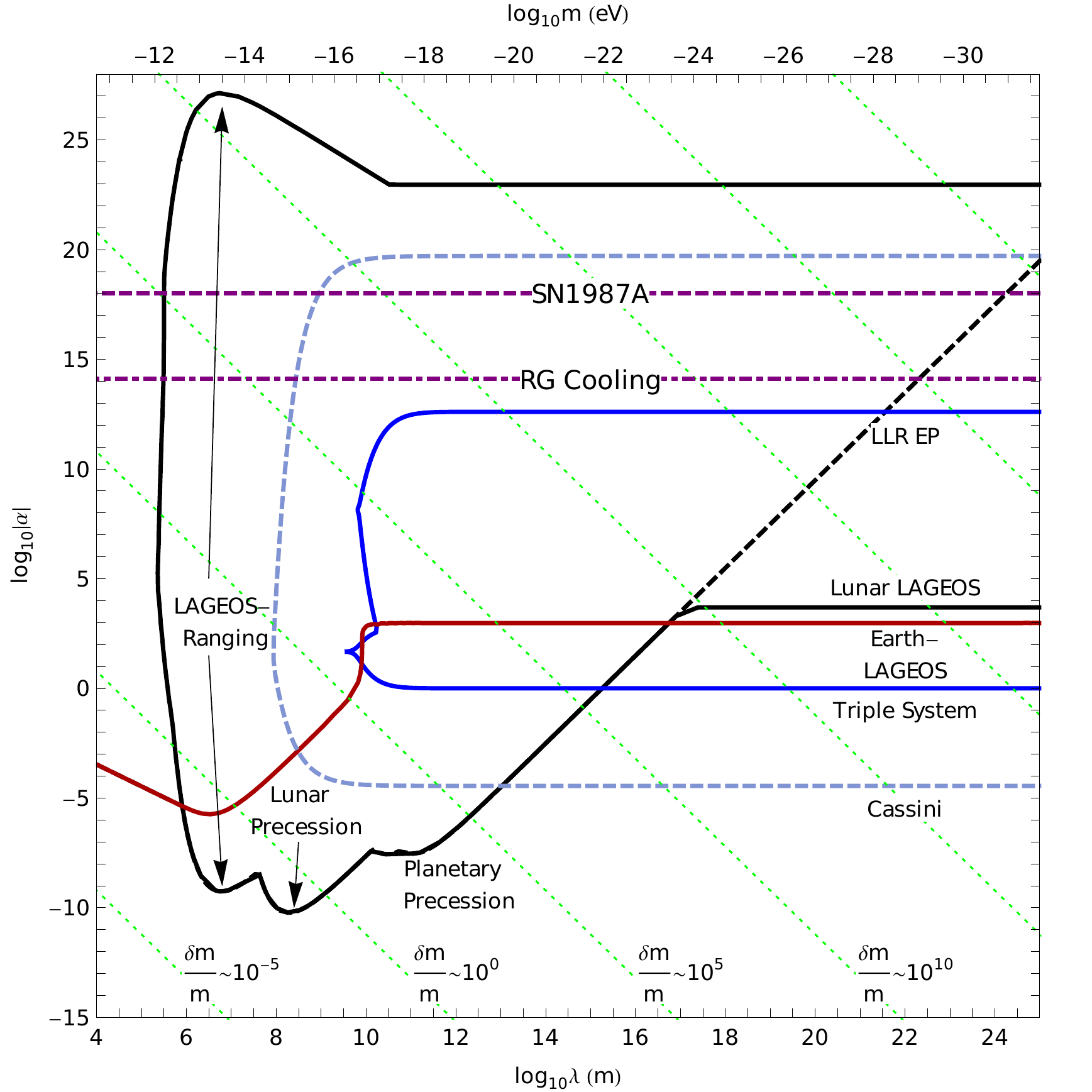}
  \caption{Constraints on fifth forces in the presence of natural quartic self-interactions from searches for EP-preserving forces excluding Earth-LAGEOS (black) and EP-violating forces (blue). The red contour indicates the constraint from Earth-LAGEOS \cite{Fischbach:1999bc}, which is unbounded above as explained in the text, thereby closing all gaps in other experiments. Searches for EP-preserving forces shown include measurements made using the LAGEOS satellite \cite{doi:10.1029/JB090iB11p09217, Fischbach:1999bc, Lucchesi:2014uza}, and the anomalous precession of the moon, Mercury and Mars \cite{Talmadge:1988qz, Dickey:1994zz, Fischbach:1999bc}. The dashed black line shows the bounds in the absence of screening, while the solid contour shows the bounds with screening. The search for light deflection by the Cassini mission \cite{Bertotti:2003rm} is shown by a dashed light blue line. Searches for EP-violating forces shown include Lunar Laser Ranging (LLR-EP)~\cite{Williams:2004qba} and the stellar triple system \triplesys \cite{Archibald:2018oxs}. These would not be present in the absence of screening. The purple dashed and dot-dashed lines indicate constraints from the cooling of SN1987A and Red Giants respectively \cite{Raffelt:1996wa,Hardy:2016kme, Knapen:2017xzo}. The dotted green contours indicate the level of tuning of the mass required as a function of $\alpha$.
    \label{fig:quartic_screened_all}}
\end{figure}

Similar figures can be made for the other experimental constraints, but ultimately they often overlap substantially, so that it is more useful to see how the combination of constraints is modified, as is shown in Fig. \ref{fig:quartic_screened_all} for a natural quartic self-interaction.

As discussed previously, inside a screened dense object sourcing the light scalar field, the field profile is roughly constant, taking on the value $\pmin$. Near the surface of the object however, the non-linearity of the potential is important, and as such, determining how experimental constraints are modified requires a numerical treatment. At distances much greater than the radius of the object, however, one is typically in a regime where the linear term dominates the equation of motion, such that we can make use of the effective charge approximation as in Refs.~\cite{Feldman:2006wg,Mota:2006fz,Mota:2006ed} -- see Eqs.~\ref{eq:effective_fifth_force} and~\ref{eq:aeff_def}. 
Experiments that employ small source/test masses (such as torsion pendulums~\cite{Adelberger:2009zz} and the MICROSCOPE satellite~\cite{Touboul:2017grn}) are not modified by screening effects at interesting values of $\alpha$ due to the smallness of technically natural self-interactions considered here. For example the critical value of $\alpha$ for the Pt/Rh test mass used in the MICROSCOPE mission is $\alpha_c \sim 10^{20}$.
Therefore, for simplicity we focus on experimental bounds coming from measurements on astronomical scales. These constraints arise from measurements involving the LAGEOS satellite \cite{doi:10.1029/JB090iB11p09217, Fischbach:1999bc, Lucchesi:2014uza}, the anomalous precession of the moon, Mercury and Mars \cite{Talmadge:1988qz, Dickey:1994zz, Fischbach:1999bc}, light deflection near the Sun~\cite{Bertotti:2003rm}, and the stellar triple system~\cite{Ransom:2014xla,Archibald:2018oxs}. 
Below, we discuss each measurement, and how it is used to constrain the effect of a fifth force. For the interested reader, a more complete treatment, including derivations of various expressions, is given in \cite{Fischbach:1999bc}.
The natural-quartic modified bounds are shown in Fig.~\ref{fig:quartic_screened_all}.


\paragraph{Earth-LAGEOS}
\label{eLAGEOS.SEC}
The Earth-LAGEOS constraint uses the quantity $\mu_\Earth(r) \equiv G(r) M_\Earth$ measured on Earth and at the LAGEOS satellite to constrain the distance variation of $G(r)$.
Using these measurements, one can extract a value for the acceleration due to Earth's gravitational pull $g_i(r)$ at the location of object $i$. 
The constraint is usually stated in terms of the $\eta$ parameter defined as
\begin{align}
\eta = \frac{g_\Earth(R_\Earth)-g_L(R_\Earth)}{g_L(R_\Earth)} \ , 
\end{align}
where $g_L(R_\Earth)$ is obtained by extrapolating the measured $g_L(r_L)$ to $R_\Earth$ using the Newtonian $1/r^2$ scaling 
  and accounting for various corrections such as deviations from perfect sphericity of the Earth~\cite{Rapp:1987}.  
We use the subscript $\Earth$ and $L$ to represent Earth and LAGEOS respectively. 
The extrapolated value is then \cite{Fischbach:1999bc}
\begin{align}
\eta = (-2 \pm 5) \times 10^{-7} \ .
\end{align}
In the Newtonian limit, the acceleration due to the Earth as measured at Earth, and as measured at LAGEOS and then extrapolated to the Earth should be identical, such that $\eta = 0$.

Because this constraint involves comparing the gravitational acceleration at the LAGEOS satellite, which is in orbit somewhat close to the Earth, with the gravitational acceleration as measured on the Earth's surface, computing the effect of screening requires additional care. When sources are well-separated, the effective charge approximation will be appropriate. However, for a measurement made on the surface of the earth, where non-linear effects are important, the effective charge approach no longer holds. At the orbit of the LAGEOS satellite, non-linear effects due to the Earth are smaller, but still important. 

A long-range force with strength $\alpha$ relative to gravity would give a non-vanishing $\eta$
\begin{align}
\eta_5 =  \left(\frac{\alpha \bar{\gamma}_\Earth(R_\Earth)  R_\Earth^2 \hspace{2pt} \mathbb{G}_\Earth(R_\Earth, m) - \alpha \bar{\gamma}_\Earth(R_L) \gamma_L R_L^2 \hspace{2pt}\mathbb{G}_\Earth(R_L, m)}{\mu_\Earth(R_L)+\alpha \bar{\gamma}_\Earth(R_L)\gamma_L R_L^2 \hspace{2pt}\mathbb{G}_\Earth(R_L, m)}\right) \ ,
\end{align}
where $\bar{\gamma}_\Earth(R_i)$ is related to the previously defined screening parameter $\gamma_\Earth$ as described below, and $\mathbb{G}_\Earth(r,m)$ is the acceleration due to to a Yukawa-type fifth force, given by
\begin{align}
\mathbb{G}_i (r,m) = G_N M_i \left(1+ m r \right) \left( \frac{e^{-m r}}{r^2}\right)F_i\left( m R_i\right) \ ,
\end{align}
where $F_i(m R_i)$ is a form factor which accounts for the extended size of the source, given by
\begin{align}
F_i(x) = \frac{3}{x^3} \left( x \cosh x - \sinh x \right) \ ,
\end{align} 
such that in the limit where $R_i \ll 1/m$, $F_i \sim 1$, and in the limit where $R_i \gg 1/m$, $F_i \sim 3e^{ m R_i}/(2 ( m R_i)^2)$ \cite{Fischbach:1999bc}.

In practice, to account for the non-linearity, we compute the derivative of the field profile $\p'(r)\vert_{r=R_\Earth,R_L}$ numerically at the Earth's surface and at LAGEOS' orbit, and compare this with the acceleration due to a pure Yukawa-type force. We parametrize the size of the screened acceleration near the surface of Earth as 
\beq
\alpha \bar{\gamma}_{\Earth}(r) \mathbb{G}_\Earth(r, m) = - \beta \p'(r), \
\eeq
where $r = R_\Earth$ or $R_L \approx 2 R_\Earth$.
We differentiate between the previously defined screening parameter $\gamma_i$ and $\bar{\gamma}_\Earth(r)$ because the latter is $r$-dependent and, crucially, exhibits different scaling as a function of $\alpha$ due to non-linear effects from being near the surface of the source. 
We can estimate the scaling of $\bar{\gamma}_\Earth$ at the Earth's surface as a function of $\alpha$ for a natural-sized quartic self-interaction term as follows~\cite{Feldman:2006wg}. 
When the equation of motion is sufficiently non-linear (i.e. when $\meff R$ is large) it can be approximated as
\beq
\p'' \sim \epsilon \p^3 - \beta \rho \theta(r-R_i) \ ,
\eeq
which can be solved to find~\cite{Feldman:2006wg}
\beq
\p \sim \pm\frac{1}{\sqrt{\frac{\epsilon}{2}}\left( r-R_i \pm \sqrt{\frac{2}{\epsilon}} \p_\text{min}^{-1}\right)}. 
\eeq
The gradient at the surface of the Earth is therefore
\beq
\p'(R_\Earth) \sim -2^{1/6} \pi^{1/3} \left(\frac{\rho_\Earth}{m_n} \right)^{2/3} \ ,
\eeq
where we have used that $\beta = \sqrt{4\pi \alpha}/\Mpl$ and $\epsilon \sim \alpha^2 (m_n/\Mpl)^4$. Therefore, for a natural-sized quartic interaction, 
\beq
\bar{\gamma}_\Earth(R_\Earth) ~ \propto ~\alpha^{-1/2} \ .
\eeq
Thus $\eta$ \textit{grows} for increasing $\alpha$, unlike the case with the other experiments we consider, where the constraints shrink as $\alpha$ grows above $\alpha_c$. The screening parameter $\bar{\gamma}_\Earth(R_L)\gamma_L$ of the acceleration at LAGEOS is also computed numerically. Because LAGEOS is orbiting at $R_L \sim 2R_\Earth$, it is in a regime where the above scaling does not apply, but neither does the estimated scaling of Eq. \ref{GammaScaling.EQ}. Instead, the scaling is intermediate, with $\bar{\gamma}_\Earth(R_L)\gamma_L \sim \alpha^{-0.6}$ obtained numerically.

The Earth-LAGEOS constraint exhibits apparent EP \emph{violation} even for an EP-preserving force, because $\bar{\gamma}_\Earth$ and $\bar{\gamma}_L$ scale differently with $\alpha$. Thus the Earth-LAGEOS bounds fifth forces with natural-sized interactions at $m\to0$, as seen in Fig. \ref{fig:quartic_screened_all} for the natural quartic interaction.

Because the self-interactions are natural-sized and therefore have $\alpha$ dependence, screening \emph{does not} result in a gap in experimental coverage opening at large $\alpha$ for the Earth-LAGEOS constraint. Thus, despite all other constraints having upper bounds in $\alpha$, Earth-LAGEOS closes off the resulting gaps.

\paragraph{Light Deflection}
Measurements of the mass of an object from gravitational 
light deflection and from orbital dynamics differ in the presence of a fifth-force, since $\p$ does not couple to light at 
the classical level (although such couplings are generated by charged matter through quantum corrections~\cite{Brax:2010uq}). 
This effect can be captured by the post-Newtonian parameter $\gamma_{\mathrm{PPN}}$.
The Cassini spacecraft has observed the gravitational deflection of radio signals as they passed close 
to the surface of the Sun and measured  $\gamma_{\mathrm{PPN}}$ to be~\cite{Bertotti:2003rm}
\beq
\gamma_{\mathrm{PPN}} - 1 = (2.1\pm2.5)\times 10^{-5}.
\eeq
We evaluate the PPN parameter following Ref.~\cite{Burrage:2017qrf}, which yields
\beq
\gamma_{\mathrm{PPN}} - 1 \approx -\frac{2\beta \p(b_{\rm min})}{\Phi_N(b_{\rm min})}
\eeq
where $b_{\min} = 1.6R_\odot$ is the minimum impact parameter of the radio signals 
sent and received by the Cassini probe~\cite{Bertotti:2003rm} and 
$\Phi_N$ is the Newtonian potential of the Sun.  The choice of $b_{\min}$ as the distance scale is not particularly important as other bounds become more important at smaller distance scales.

\paragraph{Lunar-LAGEOS}
\label{LLAGEOS.PAR}
The Lunar-LAGEOS constraint uses measurements of $\mu_\Earth(r)$ at the LAGEOS satellite and at the Moon. We use the symbol $\Earth-\Moon$ to signify the Earth-Moon distance. 
The constraint is given in terms of the parameter
\begin{align}
\eta_{LL} = \frac{\mu_\Earth(r_L) - \mu_\Earth(r_{\Earth-\Moon}) }{\left( \mu_\Earth(r_L) + \mu_\Earth(r_{\Earth-\Moon})\right)/2} \ ,
\end{align}
and is measured to be \cite{Fischbach:1999bc}
\begin{align}
\eta_{LL,\text{ meas.}} = \left( -1.8 \pm 1.6 \right) \times 10^{-8} \ .
\end{align}
Again, in the exact Newtonian limit, the expectation is that $\eta_{LL} = 0$.

In the presence of a long-range force with strength $\alpha$, $\eta_{LL}$ becomes
\begin{align}
\eta_{5,LL}  = 2 \alpha  \hspace{2pt}\gamma_\Earth  \left \{ \frac{ \gamma_L \hspace{2pt} \mathbb{G}_\Earth(R_L, m) - \gamma_\Moon \hspace{2pt} \mathbb{G}_\Earth(R_{\Earth-\Moon}, m) (R_{\Earth-\Moon}/R_L)^2}{2 \mu_\Earth / R_L^2 +\alpha \hspace{2pt} \gamma_\Earth  \left(  \gamma_L \hspace{2pt} \mathbb{G}_\Earth(R_L, m) +  \gamma_\Moon \hspace{2pt}  \mathbb{G}_\Earth(R_{\Earth-\Moon}, m) (R_{\Earth-\Moon}/R_L)^2\right) } \right\} \ ,
\label{eq:lunar_lageos_constraint}
\end{align}
where we note that the terms numerator do not have the same charges.
As a result, one finds that in the limit where $m\to 0$, $\eta_{5,LL}$ asymptotes towards a finite value:
\begin{align}
\eta_{5,LL} \sim  \alpha  \gamma_\Earth (\gamma_\Moon-\gamma_L) + \mathcal{O}(r_i^2m^2) \ ,
\end{align}
so that if $\gamma_\Moon\neq\gamma_L$, there will be effective EP violation, even if the underlying force is EP-preserving. 

\paragraph{Anomalous Precession}
\label{AnomPrec.PAR}
Bounds on anomalous precession of pericenters of objects in the Solar system provide some of the most powerful constraints on fifth forces.
We consider bounds from the LAGEOS satellite~\cite{Lucchesi:2014uza}, the Moon~\cite{Dickey:1994zz}, Mercury, and Mars~\cite{Talmadge:1988qz}.

The motion of a planet under the influence of a central force is governed by
\begin{align}
u(\theta) = u_p + u_e \cos \omega (\theta-\theta_0)  \ ,
\end{align}
where $u=1/r$. The quantity $u_p$ is related to the semi-major axis $a_p$ and the eccentricity of the orbit $\epsilon = u_e / u_p$ by $1/u_p = a_p(1- \epsilon^2)$. 
In the absence of a fifth force, $\omega = 1$ and therefore pericenter occurs at $\theta-\theta_0 = 2\pi n$, $n\in \mathbb{Z}$.
In the presence of a fifth force, $u_p$ is modified, but is still identified with the semi-major axis, so that it is not used as a constraint. 
The precession rate $\omega$ is also shifted, resulting in a pericenter shift per orbit of $\delta\theta = 2\pi \delta \omega/\omega$ where
\beq
\frac{\delta \omega}{\omega} \simeq \frac{\alpha}{2} (m a_p)^2 e^{-m a_p}.
\eeq
We note that in the presence of a Yukawa-type force, as $m\to 0$, the anomalous precession goes to zero. 
When the orbiting bodies become screened $\alpha$ is replaced by $\aeff$. If the screening is severe enough $\delta\omega/\omega$ 
  falls below the experimental limit, resulting in an upper bound on the excluded region. This was illustrated in Fig.~\ref{fig:quartic_screened_llr} for the lunar precession constraint.

The experimental constraints on the anomalous precession of the LAGEOS satellite~\cite{Lucchesi:2014uza}, the Moon~\cite{Dickey:1994zz}, Mercury and Mars~\cite{Talmadge:1988qz} are
\begin{align}
&\frac{\delta \omega}{\omega}\Big\vert_{L} = (1.4 \pm 22 \pm 270)\times10^{-13},\\ 
&\frac{\delta \omega}{\omega}\Big\vert_{\Moon} = \left( -3.0 \pm 8.0 \right)\times 10^{-12} ,\\
&\frac{\delta \omega}{\omega}\Big\vert_{\Mercury} = \left(-13 \pm 33 \right) \times 10^{-9}, \\
&\frac{\delta \omega}{\omega}\Big\vert_{\Mars} = \left(-21 \pm 29 \right) \times 10^{-9} .
\end{align}
In all cases these measurements provide the best bounds at small $\alpha$ for $1/m \sim a_p$.

\paragraph{EP Test with the Stellar Triple System}
The recently discovered stellar triple system - an inner white dwarf-pulsar binary in orbit around another white dwarf (WD) - provides 
an ideal laboratory for testing the equivalence principle~\cite{Ransom:2014xla}. 
The first tests of EP have already been carried out in Ref.~\cite{Archibald:2018oxs}, 
placing stringent bounds on violations of strong EP (SEP). 
This bound can be recast to constrain the scenario considered here since the vastly 
different sizes and densities of the white dwarves and the pulsar imply that 
their effective charges under the fifth force can be different for $\alpha\sim \mathcal{O}(1)$.

The constraint in Ref.~\cite{Archibald:2018oxs} is on the fractional difference of gravitational and inertial mass of the pulsar.
A non-zero value would lead to anomalous relative acceleration of the pulsar and inner dwarf towards the outer dwarf, leading to the bound
\beq
\eta = \frac{a_{\NS} - a_{\WD,I}}{(a_{\NS} + a_{\WD,I})/2} < 2.6 \times 10^{-6},
\label{eq:triple_system_ep_bound}
\eeq
where $a_{\NS}$ and $a_{\WD,I}$ are the accelerations of the neutron star (NS) and the inner white dwarf towards the 
outer white dwarf.
In our case, a similar anomalous acceleration arises due to different scalar charges of the inner dwarf and the neutron star.
The resulting modification to the equations of motion can be cast in a similar form to the one due to strong EP violation up 
to general relativistic corrections. Thus we apply Eq.~\ref{eq:triple_system_ep_bound} directly; a more accurate approach would be to 
simulate the orbital dynamics of the triple system as in Ref.~\cite{Archibald:2018oxs} in the presence of the $\p$-mediated 
force.

We evaluate Eq.~\ref{eq:triple_system_ep_bound} in analogy to the Lunar-LAGEOS constraint in Eq.~\ref{eq:lunar_lageos_constraint},
by computing the effective charges for the neutron star and white
dwarves as a function of the matter coupling $\alpha$. 
The masses of stars were measured to be $1.44\Msol$, $0.20\Msol$ and $0.41\Msol$ for the pulsar, inner and outer white dwarves, respectively, 
while the radius of the outer orbit is $\sim 120$ light-seconds (we neglect the small $\sim 1$ light-second radius of the binary for 
simplicity)~\cite{Ransom:2014xla}.
We use $R = 10\,\km$, $0.015R_\odot$, $0.012R_\odot$ for the radii 
of the neutron star, inner and outer white dwarves, respectively. We extract the white dwarf size from the mass-radius 
relation in Ref.~\cite{Kippenhahn:2012}. The precise values of the radii depend on the unknown stellar composition.
This uncertainty is not important because the signal arises due to the vastly different size and density
of the pulsar relative to the inner dwarf. Similarly the NS radius depends on its equation of state; 
for a neutron star of mass $1.4 M_\Sun$, the range of radii $9.9\;\mathrm{km}\lesssim R \lesssim 13.6$ km is favored~\cite{Annala:2017llu}.

\paragraph{EP Test with the Earth-Moon-Sun System}
The bound on the differential acceleration of the Earth and the Moon towards the Sun 
provides a powerful constraint on EP-violating forces~\cite{Williams:2004qba}:
\beq
\eta = \frac{|a_{\Earth} - a_{\Moon}|}{(a_{\Earth} + a_{\Moon})/2} < 1.8 \times 10^{-13}.
\label{eq:llr_ep_bound}
\eeq
This constraint is derived for the Laser Lunar Ranging data, so we refer to this limit as LLR-EP.
While the precision of Eq.~\ref{eq:llr_ep_bound} is far superior to the triple system bound in Eq.~\ref{eq:triple_system_ep_bound} and the 
relevant length scale ($\sim 1$ AU) is similar, LLR-EP is less constraining at small couplings because the 
critical values of $\alpha$ at which differential screening (and the resulting effective EP violation) 
appears are much larger than in the triple stellar system -- see Tab.~\ref{alphaCrit.TAB}.
On the other hand, this also guarantees that LLR-EP is able to constrain much \emph{larger} couplings than 
\triplesys, where the magnitude of the fifth force becomes severely screened at smaller $\alpha$ than 
for the Earth-Moon-Sun system.

\paragraph{EP Test with Binary Pulsars}
Equivalence principle-violating long-range forces can give rise to anomalous orbital eccentricity 
variation in binary systems in the field of the galaxy~\cite{Damour:1991rq}. This has been 
recently constrained using pulsar timing measurements of \textit{PSR J1713+0747}, an NS-WD binary $\sim 7$ kpc from the galactic center~\cite{Zhu:2018etc}.
While we expect this measurement to be sensitive to the $\p$-mediated force, we cannot recast it into a bound on $\alpha$ because the pulsar is immersed in DM so that the far field approximation breaks down and any screening is extremely sensitive to 
the unknown DM particle and halo properties.  

\paragraph{Scalar Radiation in Binary Systems}
Equivalence principle-violating long-range forces can also give rise to dipole gravitational radiation 
from binary star systems. Many observations have been made of systems involving a pulsar and a white dwarf companion,
such as \textit{J1738+0333}, which provides the strongest radiation constraint on EP-violating forces~\cite{Freire:2012mg, Will:2014kxa}.
However, these bounds are not competitive with the bound from the stellar triple system we consider above in our case.
This is because EP violation will only occur at the onset of screening, which in the binary system will be dictated by 
$\alpha_c$ for the pulsar. Since this will be nearly the same for both the triple system and the binary, the lower bound 
on $\alpha$ will not change. 
Additionally, because any dipole radiation is proportional to the difference in charges squared, it is further suppressed as compared to other sources.
The force range these binaries can constrain will vary depending on the system, but in general the binaries have smaller orbits
than the triple system, and will therefore only constrain parameter space that is already excluded by other searches.

\paragraph{Cooling of SN1987A, Horizontal Branch and Red Giant Stars}

Light fields with weak coupling to matter can lead to anomalous 
  cooling of stellar systems~\cite{Raffelt:1996wa}. Constraints on such interactions 
  can be obtained from the observed duration of the 
  supernova SN1987A neutrino burst, and from the agreement of 
  stellar models of horizontal branch and red giant stars with observations.
  The typical temperatures in the cores of these objects are $\sim 30\;\MeV$ for 
  the SN and $\sim 10\;\keV$ for the stars, so the resulting 
  bounds are $\p$ mass independent for $m$ considered in this work.
  The dense nuclear medium of SN1987A constrains the $\p$-nucleon coupling to be $y_N \lesssim 10^{-10}$~\cite{Hardy:2016kme, Knapen:2017xzo}. This in turn becomes a constraint on the strength of the force relative to gravity of $\alpha \lesssim 10^{17}$.
  Similarly, bounds on electron and nucleon couplings of $\p$ in red giants demand that $\alpha \lesssim 2 \times 10^{13}$~\cite{Hardy:2016kme}.

\paragraph{Environmental Dependence of Fundamental Constants}
An in-medium $\p$ shift locally modifies 
masses of the fermions that $\p$ couples to.
In dense environments like neutron stars, this can lead to an $\mathcal{O}(1)$ 
change in nucleon mass, drastically altering their properties like 
cooling and pulsation rates~\cite{Ellis:1989as}.
In the screened regime, $\p$ attains its minimum inside a neutron star; the 
resulting shift in nucleon mass is
\beq
\frac{\delta m_n}{m_n} = \beta \pmin \sim \left(\frac{\rho}{16\pi^2 m_n^4}\right)^{1/(n-1)},
\label{eq:nucleon_mass_shift_natural}
\eeq
where in the last step we plugged in the natural self-interaction, Eq.~\ref{eq:natural_self_interaction}.
Note that this is independent of the $\p$-matter coupling.
Requiring that observed pulsar periods match the expected value 
demands that~\cite{Ellis:1989as}
\beq
-1.5\lesssim \frac{\delta m_n}{m_n} \lesssim 0.6,
\label{eq:pulsar_mass_shift_bound}
\eeq
while Eq.~\ref{eq:nucleon_mass_shift_natural} gives $\delta m_n/m_n = 0.8-0.9$ for $n=3-5$.
This holds in the deep screening regime. 
The bound in Eq.~\ref{eq:pulsar_mass_shift_bound} made use of supposed period measurement of a pulsar remnant of SN1987A, which has not been confirmed by later observations~\cite{Zhang:2018dez}. However, we expect that similar results can be obtained for other systems. We do not show this constraint in Fig.~\ref{fig:quartic_screened_all} since 
it depends sensitively on the neutron star equation of state and on the transition between screened and unscreened regimes.

Additional strong constraints on shifts in fundamental constants can be obtained from cosmology \cite{Brax:2004qh,Mota:2006fz, Olive:2007aj, Uzan:2010pm, Coc:2012xk}. In particular, nucleon-nucleon interactions during Big Bang Nucleosynthesis are sensitive to shifts in masses and in the strong coupling constant. These constraints are stronger than those from pulsar periods described above, but are model-dependent, so we do not plot them here. 

\section{Higher-Dimensional Operators} \label{sec:higherdim}

Higher dimensional operators provide a unique opportunity for experimentalists to test assumptions that are typically held about quantum gravity. To see how this works in an explicit example, consider the dimension 5 operator so that the leading order equation of motion is
\bea
\nabla^2 \p = \frac{\p^{4}}{\Lambda}.
\eea
Imagine that a new force sourced by the Sun is discovered, with a strength slightly below the current bounds, and that it has the standard $1/r$ scaling with distance. Because we are assuming that a $\p \sim Q/r$ potential has been measured, the quintic operator must not cause deviations in the $1/r$ behavior of the potential. 
Roughly speaking, this is satisfied if
\bea
\frac{1}{r^2} \frac{Q}{r} \gtrsim \frac{1}{\Lambda} \frac{Q^4}{r^4},
\eea
i.e. if the derivative terms are dictating the falloff of the potential and not the quintic interaction.  Plugging in the current constraints on $Q_{\Sun} \lesssim 10^{35}$ at distances $ r \sim 10^{11}$ m, we see that this measurement would place the constraint
\bea
\Lambda \gtrsim 10^{60} \Mpl.
\eea
Note that this strong constraint is placed despite the fact that actual field value of the fifth force, $\p \sim 10^{-10} \Mpl$, is much smaller than the Planck scale.
Now the ``natural" size of this quintic piece is $\Lambda \sim m_f/y^5 \gg 10^{60} \Mpl$ which satisfies the constraint (see Eq.~\ref{eq:natural_self_interaction}) for $\alpha=1$. The ``natural" size of a gravitational contribution to this piece would also be suppressed, $\Lambda \sim \Mpl/y^5$ so that as $y \rightarrow 0$, the scalar becomes massless.
However, quantum gravity has not been experimentally tested. It is plausible that standard field theory rules break down and quantum gravity generates corrections with $\Lambda \sim \Mpl$, i.e. gravity for some reason does not respect the shift symmetry of the scalar. This is expected since general arguments imply that quantum gravity does not respect global symmetries~\cite{Kamionkowski:1992ax,Holman:1992va}.
\emph{Thus, a discovery of a scalar fifth force and a confirmation of its $1/r$ potential would place the first ever constraints on the behavior of quantum gravity.} In particular, the existence of such a fifth force would suggest that quantum gravity does indeed respect shift symmetries.

The previous example shows that if a fifth force were ever discovered, the form of its potential would constitute the first experimental constraints on quantum gravity. Repeating the above gedanken experiment with an arbitrary higher dimensional operator 
\bea
\nabla^2 \p = \frac{\p^{n-1}}{\Lambda^{n-4}}
\eea
we find that measurement of a $\p = Q/r$ potential at distance $r$ from the source would place the constraint
\bea
\Lambda > \frac{Q^{\frac{n-2}{n-4}}}{r}.
\eea
For $n=5$ and $n=6$, very strong super-Planckian constraints can be placed, while for $n \geq 7$, the constraints are all sub-Planckian.

\paragraph{A more detailed analysis}

\begin{figure}
  \centering
  \includegraphics[width=10cm]{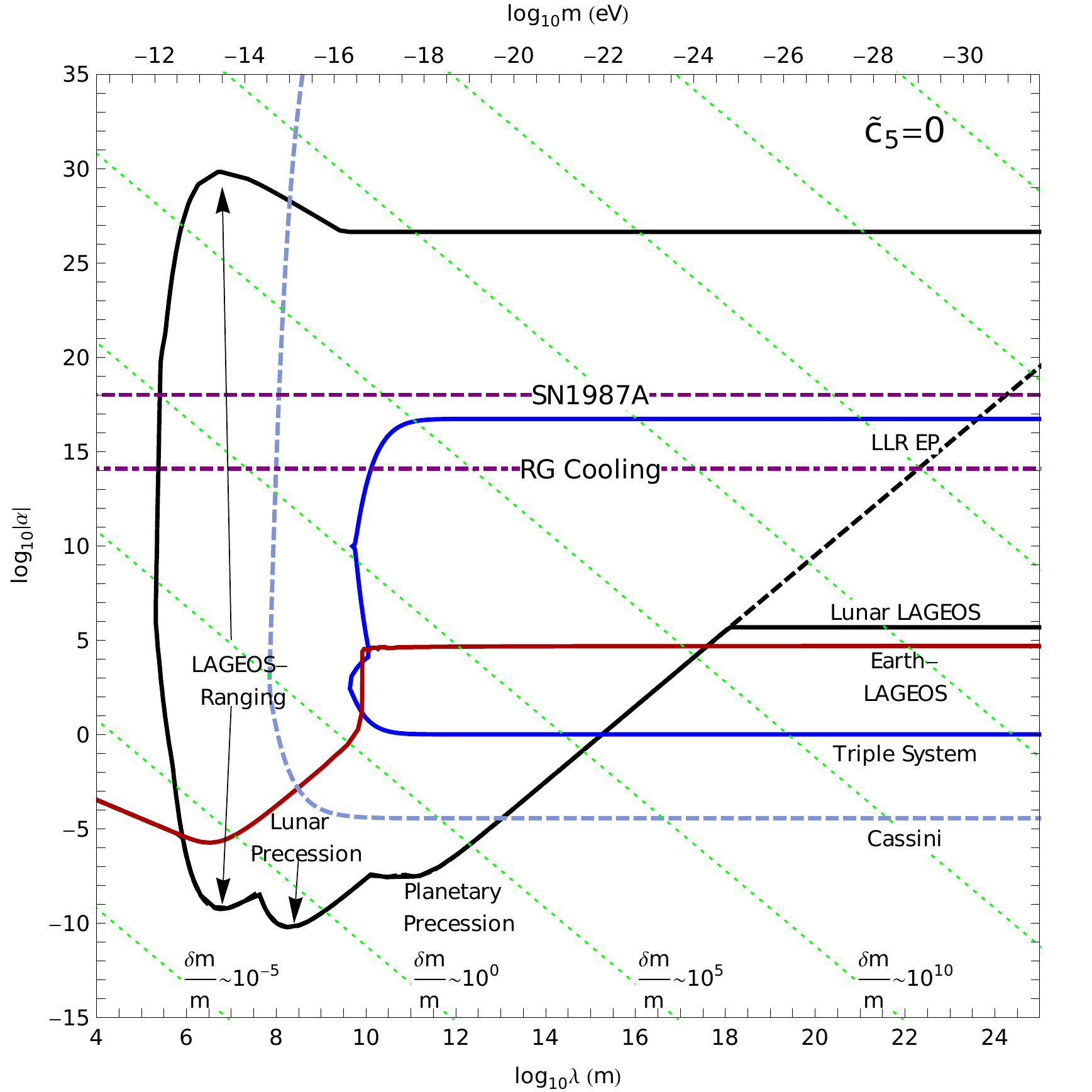}
  \caption{Constraints on fifth forces in the presence of natural quintic self-interactions from searches for EP-preserving forces excluding Earth-LAGEOS (black) and EP-violating forces (blue). The red contour indicates the constraint from Earth-LAGEOS \cite{Fischbach:1999bc}, which is unbounded above as explained in the text, thereby closing all gaps in other experiments. Searches for EP-preserving forces shown include measurements made using the LAGEOS satellite \cite{doi:10.1029/JB090iB11p09217, Fischbach:1999bc, Lucchesi:2014uza}, and the anomalous precession of the moon, Mercury and Mars \cite{Talmadge:1988qz, Dickey:1994zz, Fischbach:1999bc}. The dashed black line shows the bounds in the absence of screening, while the solid contour shows the bounds with screening. The search for light deflection by the Cassini mission \cite{Bertotti:2003rm} is shown by a dashed light blue line. Searches for EP-violating forces shown include Lunar Laser Ranging (LLR-EP) \cite{Williams:2004qba} and the stellar triple system \triplesys \cite{Archibald:2018oxs}. These would not be present in the absence of screening. The purple dashed and dot-dashed lines indicate constraints from the cooling of SN1987A and Red Giants respectively \cite{Raffelt:1996wa,Hardy:2016kme, Knapen:2017xzo}. The dotted green contours indicate the level of tuning of the mass required as a function of $\alpha$.
    \label{fig:quintic_screened_all}}
\end{figure}

Another way to interpret the discussion of Planck scale corrections is in terms of the impact of a tree-level contribution to the dimension-5 operator. The natural radiatively-generated size of the coefficient in front of the dimension-5 operator is 
\beq
\frac{c_5}{\Lambda} \sim \frac{(4\pi\alpha)^{5/2}}{16\pi^2} \left(\frac{m_n}{\Mpl}\right)^4 \frac{1}{\Mpl}
\sim 10^{-76}\frac{\alpha^{5/2}}{\Mpl}.
\eeq
The natural dimension 5 operators lead to screening in various sources when $\alpha$ exceeds the critical 
values listed in Tab.~\ref{alphaCrit.TAB}. 
The smallness of this coefficient means that $\alpha$ must be larger 
to initiate screening, relative to the quartic self-interactions considered in Sec.~\ref{sec:quartic_and_higher}.
As a result, the natural quintic self-interaction does not open any previously-excluded parameter space as 
we show in Fig.~\ref{fig:quintic_screened_all}. In fact, the non-linearity implies additional constraints 
from the pulsar triple and the Moon-LAGEOS-Sun systems due to effective EP-violation, as described in Sec.~\ref{sec:quartic_and_higher}.

\begin{figure}[t]
\centering
\includegraphics[width=300pt]{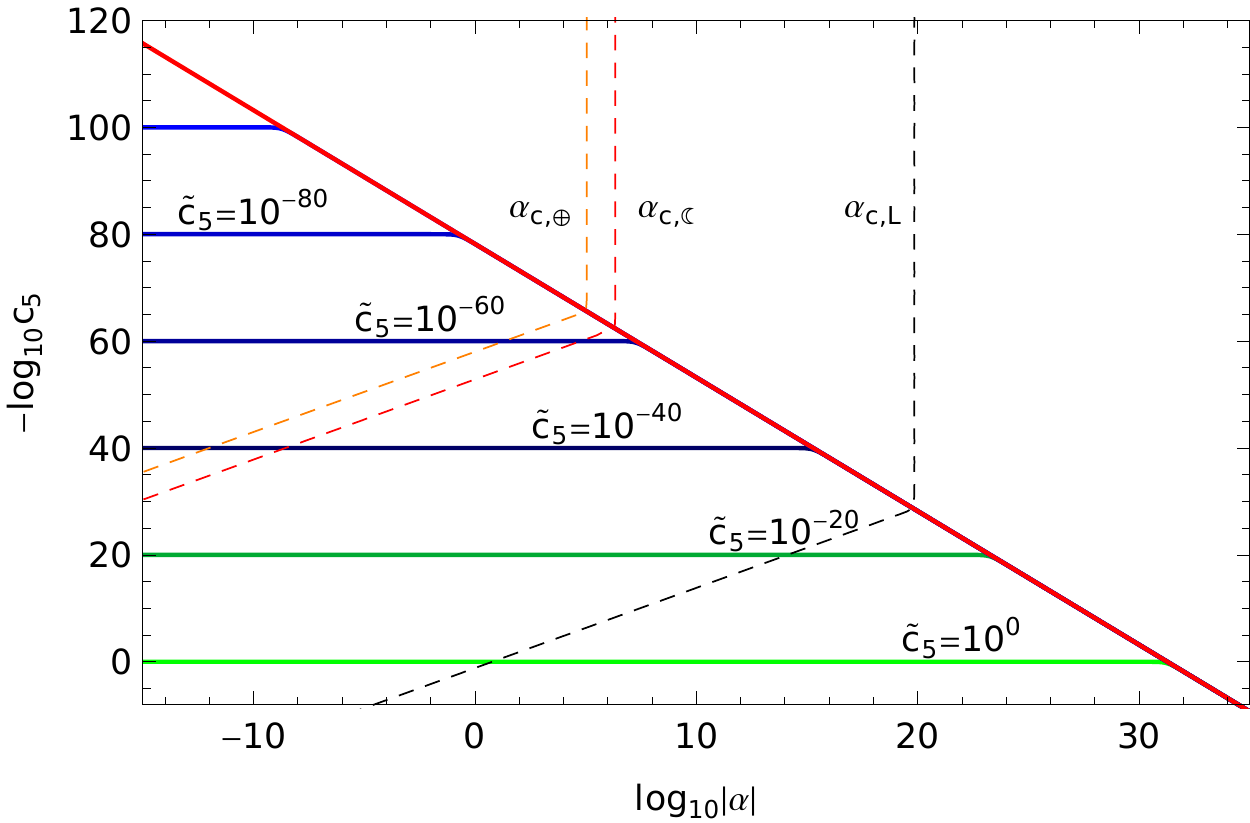}
\caption{Natural value of the quintic dimension-5 operator coefficient $c_5/\Lambda$ (red, normalised to $\Lambda=\Mpl$), as a function of $\log_{10}|\alpha|$. Possible UV tree-level contributions (green/blue contours) $\tilde{c}_5$ are independent of $\alpha$.
Also shown as dashed red, orange and black lines are contours of $\alpha_c^{(5)}$ for the Moon, Earth and LAGEOS respectively.
}
\label{d5Coefficient.FIG}
\end{figure}

However, this story qualitatively changes if there is a tree-level contribution to the dimension-5 operator which is not generated by the matter-induced radiative corrections. 
Such a contribution could, for example, arise due to quantum gravity effects as described above. 
Then, we should write
\beq
\frac{c_5}{\Lambda} \sim \frac{(4\pi \alpha)^{5/2}}{16\pi^2} \left(\frac{m_n}{M_{\text{Pl}}}\right)^4 \frac{1}{\Mpl}  + \frac{\tilde{c}_5}{\Mpl}\ ,
\eeq
where $\tilde{c}_5$ is some coefficient which is not calculable in the infrared (IR).
The tree-level contribution dominates over the natural radiative correction when
\beq
\tilde{c}_5 \gtrsim 10^{-76}\alpha^{5/2}.
\eeq
This can be seen in Fig. \ref{d5Coefficient.FIG}, where we show contours of $c_5$, with $\Lambda$ normalised to $\Mpl$, for different values of the tree-level UV contribution, $\tilde{c}_5$.

A consequence of this transition from natural IR to unknown UV contribution is that the critical value of the force strength $\alpha_c^{(5)}$ will change for all objects. The critical force strength remains independent of the mass of the light scalar, but it depends on $\tilde{c}_5$, as seen also in Fig. \ref{d5Coefficient.FIG}.

Thus, if there is a tree-level contribution in addition to the natural radiative correction term in the higher-dimensional operator, the bounds can change quite drastically. From studying Fig. \ref{d5Coefficient.FIG}, we can for example estimate the values of $\tilde{c}_5$ for which the anomalous lunar precession constraint will change substantially. We see that for $\tilde{c}_5 \gtrsim 10^{-65}$, $\alpha_{c,\Earth}^{(5)}$ begins to decrease, and similarly for the moon for $\tilde{c}_5 \gtrsim 10^{-62}$. At its strongest point, the anomalous lunar precession constraint is $|\alpha| \lesssim 10^{-10}$. Therefore, if at some point $\alpha_{c,i}^{(5)}$ drops below this value, there is no constraint at all, because $\alpha_\text{eff} \lesssim 10^{-10}$ everywhere. We see that this condition is achieved for $\tilde{c}_5 \gtrsim 10^{-40}$. Thus, we can break down the anomalous precession constraint into three regimes
\begin{itemize}
\item $\tilde{c}_5 \lesssim 10^{-60}$: \hspace{0.2 cm} the constraint is only cut off above the natural values of $\alpha_{c,i}$ as given in Table \ref{alphaCrit.TAB}.
\item $10^{-40} \gtrsim \tilde{c}_5 \gtrsim 10^{-60}$: \hspace{0.2 cm} the constraint remains cut off above natural values of $\alpha_{c,i}$, but screening begins at lower values of $\alpha$ due to the true $\alpha_{c,i}$ being smaller.
\item $\tilde{c}_5 \gtrsim 10^{-40}$: \hspace{0.2 cm} there is no constraint.
\end{itemize}

\begin{figure}[t]
\centering
\includegraphics[width=300pt]{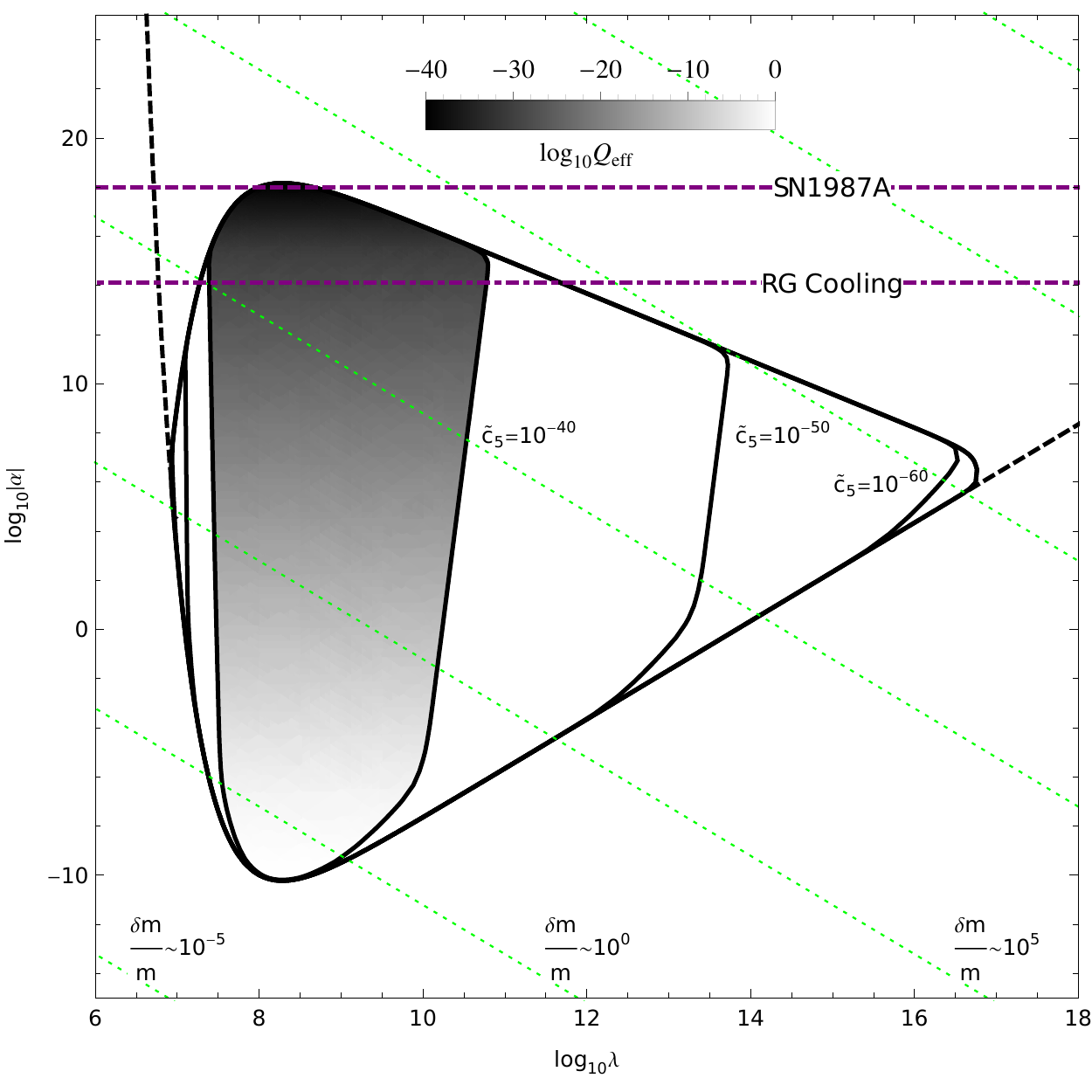}
\caption{
  Impact of a tree-level contribution $\tilde{c}_5/\Mpl$ to the quintic self-interaction 
on the anomalous lunar precession bound. As $\tilde{c}_5$ is increased from $10^{-60}$ and $10^{-40}$ 
the constrained region shrinks because the onset of screening occurs at smaller and smaller $\alpha_c$.
}
\label{d5LLR.FIG}
\end{figure}

This is illustrated in Fig. \ref{d5LLR.FIG} for the anomalous lunar precession bound, where as we vary $\tilde{c}_5$, the constrained region shrinks as we go between $10^{-60} \lesssim \tilde{c}_5 \lesssim 10^{-40}$, and would disappear if we plotted $\tilde{c}_5\gtrsim 10^{-40}$.

\begin{figure}[t]
\centering
\includegraphics[width=300pt]{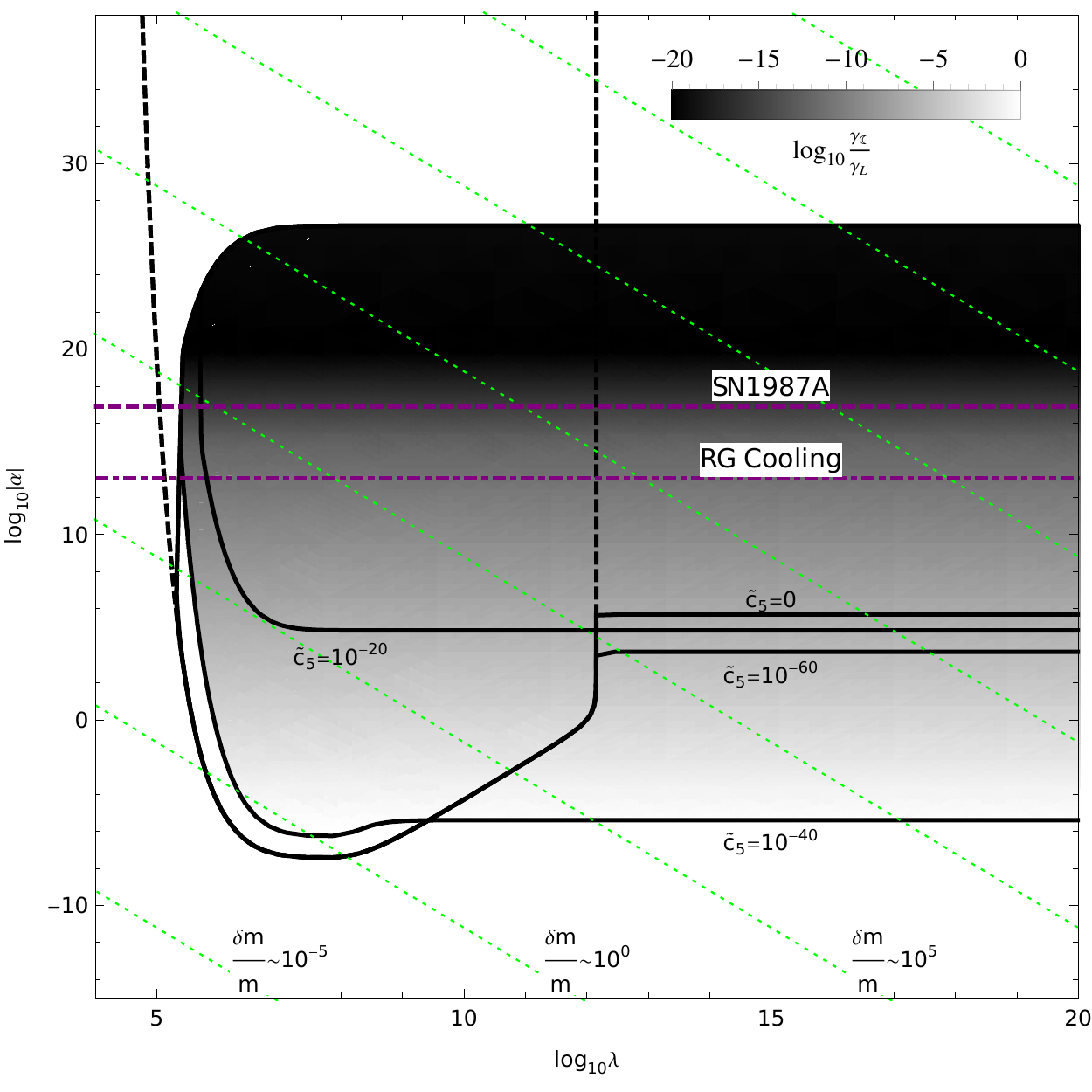}
\caption{Variation of the Lunar-LAGEOS constraint for $d=5$ self-interactions 
  with a tree-level contribution $\tilde{c_5}/\Mpl$ to the Wilson coefficient as we change 
  $\tilde{c}_5$ between $10^{-60}$ and $10^{-20}$. At first, the excluded region grows 
  are large $\lambda$ but then starts to decrease when $\tilde{c}_5$ is large enough.
  The non-monotonic scaling of the bound is explained in the text.}
\label{d5LunarLAGEOS.FIG}
\end{figure}

Similar considerations apply to the other anomalous precession constraints. For the more complex Lunar-LAGEOS and Earth-LAGEOS constraints, the behaviour is similar, but the implication of a large tree-level UV contribution can be more interesting. Now, since $\alpha_{c,\Moon}$ and $\alpha_{c,\Earth}$ decrease as we increase $\tilde{c}_5$, the onset of the effective EP violation we remarked on earlier begins at smaller values of $\alpha$, expanding the excluded region at large $\lambda$. However, for the values of $\tilde{c}_5$ for which the constraint does not disappear, the LAGEOS $\alpha_c$ is not modified, 
so the large-$\alpha$ upper bound of the excluded region is unchanged (the upper bound occurs when \emph{all} objects are screened). When $\tilde{c}_5$ is increased further, screening begins to weaken bounds in the entire 
mass range. This behavior is shown in Fig. \ref{d5LunarLAGEOS.FIG} for Lunar-LAGEOS.
Thus, as seen in Fig. \ref{d5LunarLAGEOS.FIG} for certain values of $c_5$, the constrained region \textit{grows}, while for others it shrinks. 
The non-monotonic scaling of the experimental constraint can be simply understood from 
its large $\lambda$ limit ( $1/m \gg r_{\Earth-\Moon}$) which can be written as
\beq
\alpha \lesssim \frac{\eta}{\gamma_\Earth \left( \gamma_\Moon - \gamma_L \right)} \ ,
\eeq
where we kept only the linear term in $\eta \lesssim\mathcal{O}(10^{-8})$.
In the situation where $\tilde{c}_5$ dominates $c_5$, $\gamma_i$ scales as $\alpha^{-1/2}$ (see Eq.~\ref{QeffScaling.EQ}). 
As $\tilde{c}_5$ is increased as in Fig.~\ref{d5LunarLAGEOS.FIG}, $(\gamma_\Moon - \gamma_L) \rightarrow -1$ because LAGEOS is not screened ($\gamma_L \approx 1$), while $\gamma_\Moon$ quickly becomes small. This enhances the effective EP violation.
In this limit we can rewrite the above constraint as $\alpha \lesssim \eta /(\alpha_{c,\Earth}/\alpha)^{1/2}$. 
Thus, as $\tilde{c}_5$ is increased further, $\alpha_{c,\Earth}$ decreases and the overall bound on $\alpha$ weakens.

For Earth-LAGEOS, a similar estimate can be made. In the $m\to0$ limit, the constraint on $\alpha$ can be approximated as 
\beq
\alpha \lesssim \frac{\eta}{\bar{\gamma}_\Earth(R_\Earth) - \bar{\gamma}_\Earth(R_L)} \ ,
\eeq
with $\bar{\gamma}_i$ defined as in Sec. \ref{eLAGEOS.SEC}. Once one is in the screening regime above $\alpha_c$, the scaling of the parameters 
is $\bar{\gamma}_\Earth(R_\Earth) \sim(\alpha_{c,\Earth}/\alpha)^{0.5}$, and $\bar{\gamma}_\Earth (R_L) \sim (\alpha_{c,\Earth}/\alpha)^{0.4}$. 
Therefore $\eta$ grows rapidly with $\alpha$ above $\alpha_c$, so that the Earth-LAGEOS bound tracks $\alpha_{c,\Earth}$ as in Fig. \ref{d5Coefficient.FIG}.

\section{Cubic Interactions \label{sec:cubic}}

The cubic self-interaction is qualitatively different from the 
quartic and higher interactions in that it is a relevant 
operator at low energies. Thus, it modifies 
the large distance behaviour of $\p$ and the 
force it mediates. 
We discuss these two aspects in greater detail in Sections~\ref{CubicForce.SEC} and~\ref{VacuumDecay.SEC} below.

In this section we consider the potential
\bea
\label{eq: cubic}
V = \frac{1}{2}m^2 \p^2 + \frac{1}{3}\kappa \p^3 - \beta \rho \p \ ,
\eea
and assume that there is a small quartic term for stability as $\p \rightarrow \pm \infty$.
Imagine a source with charge $Q$ that gives a standard $\p = Q/r$ potential close to the source. 
Relevant operators modify the behavior of the potential 
after some characteristic cross-over distance. 
For example, in the case of a mass term, the $Q/r$ potential becomes manifestly Yukawa-like at a distance $r \sim 1/m$.
The cross-over occurs when the gradient terms become comparable to 
the potential terms. 
We can therefore estimate the cubic term cross-over distance $r_c$ (assuming $\kappa \gg m$) as
\bea
\frac{1}{r_c^2} \frac{Q}{r_c} \sim \kappa \frac{Q^2}{r_c^2} \Rightarrow r_c \sim \frac{1}{\kappa Q}.
\eea
Unlike the case of the mass term, the critical radius at which the cubic takes over depends on 
the source charge $Q$.
Solving for the falloff of $\p$ at large radii we find
\bea
r < r_c: &\qquad& \p \sim \frac{Q}{r} \\
r > r_c: &\qquad& \p \sim \frac{1}{\kappa r^2}. 
\eea
This scaling continues until $r\sim 1/m$ when $\p$ enters the linear regime and the field profile becomes Yukawa-like. If $Q \kappa < m$ then this occurs before the $\p^3$ has a chance to dominate.

A $\p^3$ interaction has two important differences from a quartic. The first difference is that even if there is no screening or enhancement taking place, $\kappa$ modifies the long distance behavior of $\p$ sourced by an object~\cite{Sanctuary:2008jv}. Thus, a different analysis/reinterpretation of data will be needed to constrain this new fall off. 
The modification of the fifth force in the presence of a cubic self-interaction term is discussed in Section \ref{CubicForce.SEC}.

The second difference is that a $\p^3$ interaction is unbounded from below as $\p\rightarrow \pm \infty$, depending on the sign of $\kappa$. Having an unbounded potential means that there are new constraints coming from vacuum decay, as well as the possibility of enhancement playing a role. If $\beta \kappa > 0$, then finite density effects will screen forces, leading to phenomenology similar to that considered 
  in Sections~\ref{sec:quartic_and_higher} and~\ref{sec:higherdim}. We consider this possibility in Sec.~\ref{CubicForce.SEC}. If, on the other hand, $\beta \kappa < 0$, finite density effects will anti-screen (enhance) forces.  Interestingly, the strongest constraints on the case where forces are enhanced, actually come from induced vacuum decay of the universe rather than fifth force experiments. These constraints are discussed in Sec.~\ref{VacuumDecay.SEC}.

\begin{figure}
  \centering
  \includegraphics[width=10cm]{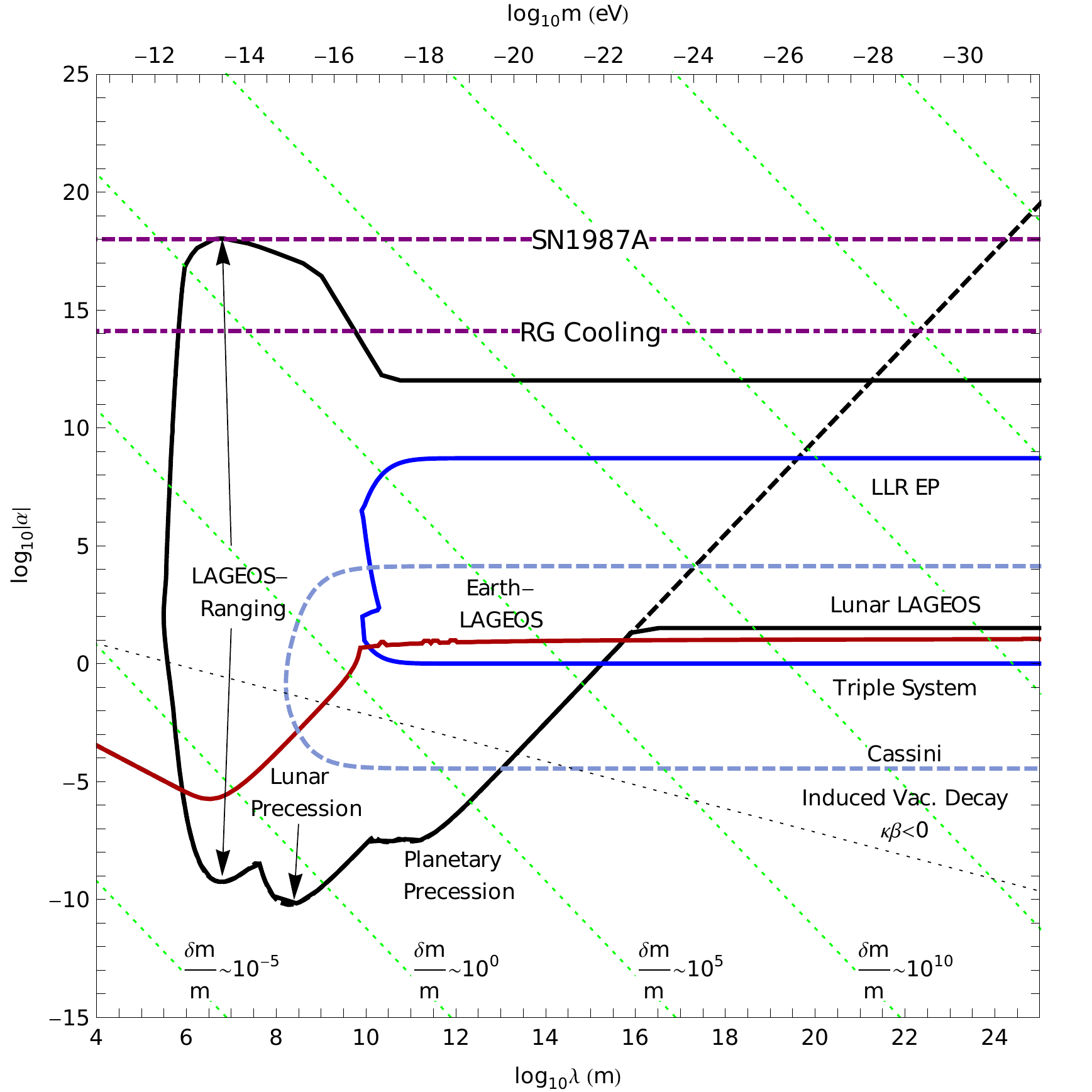}
  \caption{Constraints on fifth forces in the presence of natural cubic self-interactions from searches for EP-preserving forces excluding Earth-LAGEOS (black) and EP-violating forces (blue). The red contour indicates the constraint from Earth-LAGEOS \cite{Fischbach:1999bc}, which is unbounded above as explained in the text, thereby closing all gaps in other experiments. Searches for EP-preserving forces shown include measurements made using the LAGEOS satellite \cite{doi:10.1029/JB090iB11p09217, Fischbach:1999bc, Lucchesi:2014uza}, and the anomalous precession of the moon, Mercury and Mars \cite{Talmadge:1988qz, Dickey:1994zz, Fischbach:1999bc}. The dashed black line shows the bounds in the absence of screening, while the solid contour shows the bounds with screening. The search for light deflection by the Cassini mission \cite{Bertotti:2003rm} is shown by a dashed light blue line. Searches for EP-violating forces shown include LLR-EP \cite{Williams:2004qba} and the stellar triple system \triplesys \cite{Archibald:2018oxs}. These would not be present in the absence of screening. The purple dashed and dot-dashed lines indicate constraints from the cooling of SN1987A and Red Giants respectively \cite{Raffelt:1996wa,Hardy:2016kme, Knapen:2017xzo}. The dotted black line shows the constraint for $\kappa\beta<0$ where vacuum decay would be induced in Neutron Stars. The dotted green contours indicate the level of tuning of the mass required as a function of $\alpha$.
  \label{fig:cubic_screened_all}}
\end{figure}

\subsection{Screening and Modified Fifth Forces}
\label{CubicForce.SEC}
When the scalar has cubic self-interactions, near the source, the modified gravitational force of Eq. \ref{eq:standard_fifth_force} is not applicable, since the fifth force potential is not Yukawa-like as it is for quartic and higher self-interactions. We account for this by modifying the gravitational potential to read
\begin{align}
V_{5, ij}(r, m) = \frac{G_N M_i M_j }{r}\left(1 + \alpha \hspace{2pt} \gamma_i \hspace{2 pt} \gamma_j \hspace{2pt} e^{-mr}\left( 1 + \frac{f(\kappa)}{r}\right)\right) \ ,
\label{eq:cubic_potential_guess}
\end{align}
where the coefficient $f(\kappa)$ has dimension $[ M^{-1}]$, and encodes the dependence on the dimensionful trilinear coupling $\kappa$. 
This function should have the limit $f(\kappa)\to0$ as $\kappa\to0$, so that one recovers the Yukawa-type potential in that limit, and behaves as $f(\kappa) \sim \kappa^{-1}$ in the large $\kappa$ limit. 
In our numerics, we fit for $f(\kappa)$ and find the expected behaviour in these limits.

While the functional form in Eq.~\ref{eq:cubic_potential_guess} does not solve the EOM (there is no analytic solution), we 
  find that it provides an adequate fit to the numerical solutions, so we use it to estimate the 
experimental bounds below. We also note that when the cubic term dominates the evolution of $\p$ far 
from sources (but at $r<1/m$), the superposition principle fails since the EOM is non-linear in 
the field overlap region between two sources. Therefore, the recasting of the experimental 
limits for the cubic self-interaction is approximate at best; a correct treatment would 
involve solving the full three-dimensional EOM in the presence of multiple sources for 
each system, which is beyond the scope of this work. 
We note, however, that many constraints are strongest at $1/m \sim \lambda_{\rm exp}$ where 
$\lambda_{\rm exp}$ is the relevant experimental length scale and so $\p$ begins 
to enter the linear regime, where our calculations are reliable. The other 
important bounds, Earth-LAGEOS, the Cassini light deflection measurement and induced vacuum decay (discussed in the next section), 
do not rely on superposition.

With the above caveats in mind, we can compute the acceleration due to the potential in Eq.~\ref{eq:cubic_potential_guess}: 
\begin{align}
\mathbb{G}_i (r,m) = G_N M_i  \left(1 + mr+\frac{f(\kappa)}{r} \left(2+ m r\right)\right)\left(\frac{e^{-m r}}{r^2}\right) \tilde{F}_i\left( m \hspace{2pt}R_i\right) \ ,
\end{align}
where $\tilde{F}_i\left( m \hspace{2pt}R_i\right)$ is a form factor accounting for the extended size of the object, similar to the one obtained for a Yukawa-like potential described in Section \ref{LLAGEOS.PAR}.

Given this modified gravitational acceleration, we can then compute how all the constraints described in Section \ref{sec:quartic_and_higher} apply to the case of cubic self-interactions. The only constraint that changes qualitatively is the anomalous precession of celestial bodies. Repeating the analysis in Section \ref{AnomPrec.PAR}, we find
\beq
\frac{\delta \omega}{\omega}  \simeq \frac{\alpha}{2}e^{-m a_p} \left( (m a_p)^2 + a_p m^2 f(\kappa) + 2 m f(\kappa) + \frac{2 f(\kappa)}{a_p} \right) \ . 
\eeq
We see that now $\delta \omega/\omega$ asymptotes to $\alpha f(\kappa) /a_p$ in the $m\to0$ limit, as opposed to $0$ for a Yukawa-type force. However, in practice because $f(\kappa)\to 0$ as $\kappa\to0$, and $f(\kappa)\sim \kappa^{-1}$ as $\kappa\to\infty$, when we solve numerically for $f(\kappa)$ to compute the cubic constraints shown in Fig. \ref{fig:cubic_screened_all}, we find that the $m\to0$ behaviour of the constraints does not change with respect to quartic and higher self-interactions.

In Fig. \ref{fig:cubic_screened_all} we fix $\kappa$ to its natural lower bound 
\beq
\kappa \sim \frac{(4\pi \alpha)^{3/2}m_n}{16\pi^2} \left(\frac{m_n}{\Mpl}\right)^3,
\eeq
and estimate the resulting constraints as described above. As $\alpha$ is increased the onset of screening occurs earlier, 
because the natural cubic interaction is parametrically larger than the quartic and quintic terms considered before (see Tab.~\ref{alphaCrit.TAB}).
While bounds from certain measurements are cut off at large coupling (and below the stellar cooling bounds) as a result of this screening, we see that the Earth-LAGEOS constraint excludes these regions as in Sections~\ref{sec:quartic_and_higher} and~\ref{sec:higherdim}.

\subsection{Enhancement, and Spontaneous and Induced Vacuum Decay}
\label{VacuumDecay.SEC}

A theory with a large cubic coupling and small mass and quartic coupling is necessarily in a metastable vacuum. Thus, tunneling to a deeper minimum of the potential can occur.
The bounce action determining the false vacuum lifetime for such a theory is~\cite{Adams:1993zs}
\beq
S_E \approx \frac{205 m^2}{\kappa^2}.
\eeq
Requiring that the metastable vacuum is longer lived than the age of the universe places the rough constraint that
\beq
\kappa < \mathcal{O}(1)\; m,
\eeq
where the uncertainty on the $\mathcal{O}(1)$ coefficient comes from estimating the 
pre-exponential factor in the decay rate.
For the natural cubic in Eq.~\ref{eq:natural_self_interaction}, this bound translates into a constraint on $\alpha$:
\beq
\alpha < 10^{25} \left(\frac{10^{6}\,\mathrm{m}}{\lambda}\right)^{2/3},
\eeq
where $\lambda = 1/m$ is the vacuum range of the force.

The \emph{spontaneous} decay of the vacuum places a relatively weak constraint that is surpassed by 
various astrophysical tests for all but the smallest of masses.  This constraint is present for both $\beta \kappa >0$
and $\beta \kappa < 0$, and is the dominant constraint for small masses.

When $\beta \kappa < 0$, a much stronger constraint can be obtained from \emph{induced} vacuum decay. 
A dense, macroscopic object modifies the $\p$ potential in a spacial volume. If this source is dense enough,
it can catalyze the creation of a bubble of true vacuum.
Interestingly, the induced vacuum decay becomes important much earlier than when $1/m_\text{eff} \sim R$. 
This can be seen by noting that in the limit where the vacuum mass is small, any perturbation can cause $\p$ to start its inevitable descent to the true minimum.

To see how the decay of the vacuum can be induced, we first consider a region of space of radius $R_0$ over which $\p$ has been pushed to some field value $\p_0 > 0$, assuming that $\beta >0$ and $\kappa < 0$ so that enhancement will occur.  If $\p_0 \gtrsim |m^2/\kappa|$, then the field is probing a region of field space where the average energy density is negative and thus wants to expand. 
This is counter-acted by the gradient energy in the bubble wall (the transition region between $\p_0$ inside and $\p\approx 0$ outside), which favors contraction.

Numerically, we will be interested in the limit where $R_0$ is small so that near this region of space, gradient energy dominates and pulls it towards the origin in a standard
\bea
\p \sim \frac{\p_0 R_0}{R}
\eea
form.  Eventually, the relevant operator $\kappa \p^3$ will be comparable to the gradient energy at a critical radius $R_c$
\bea
|\kappa| \p(R_c)^3 \sim \frac{\p(R_c)^2}{R_c^2} \Rightarrow R_c \sim \frac{1}{\p_0 R_0 | \kappa |}.
\eea
At this point the cubic will cause a runaway towards the true minimum of the potential, unless 
the mass term is more important and drags it towards $\p = 0$.
The requirement that the mass term is more important at $R_c$ can be stated in two equivalent ways.
The first is that $R_c > 1/m$ so that the evolution of $\p$ for $R > R_c$ is never dominated by the cubic, but instead becomes 
Yukawa-like. The second equivalent way of phrasing this is to require that $\p(R_c) \lesssim m^2/\kappa$ so that $\p$ is in the basin of attraction of the origin once the cubic term becomes important in the evolution of $\p$ in $R$. 
A runaway would invalidate our starting assumption that we live in the vacuum at $\p = 0$ with a small mass, so that we arrive at the constraint
\bea
\label{Eq: cubic general}
|\kappa| < \frac{m}{\p_0 R_0}.
\eea
This constraint has the form that we would expect. A region of space with a large $\p_0$ and/or large $R_0$ would destabilize the space more easily while a large $m$ prevents destabilization. As $m \rightarrow 0$, the minimum near the origin disappears and any perturbation would cause the entire universe to roll away from the inflection point.

Note that Eq.~\ref{Eq: cubic general} is conservative.  It is requiring that there at least exists a stable field configuration which falls off towards $\p = 0$ at infinity.  An actual physical system due to its evolution may not actually end up in this field configuration and might cause induced decay of the universe earlier.

Let us now check if the densest objects in the universe, neutron stars, will create a region of space that satisfies this criteria.  Near the location where the exclusions are, neutron stars source $\p$ without any enhancement type effects.  In this case, the $\p$ at the surface of the neutron star is
\bea
\p_{\NS} \sim \frac{\beta M_{\NS}}{R_{\NS}},
\eea
where we identified $R_0 = R_{\NS}$ and $\p_0 = \p_{\NS}$.
Requiring that this initial perturbation caused by the neutron star does not grow and using the constraint 
in Eq.~\ref{Eq: cubic general} gives
\bea
|\kappa| < \frac{m}{\beta M_{\NS}}.
\eea
For the natural cubic in Eq.~\ref{eq:natural_self_interaction}, this bound translates into a constraint on $\alpha$:
\beq
\alpha < 0.7 \left(\frac{10^6\;\mathrm{m}}{\lambda}\right)^{1/2},
\eeq
where we took $M_{\NS} \approx 1.4\Msol$.
The strength of this constraint can be seen in Fig. \ref{fig:cubic_screened_all}.
For small masses $m \lesssim 10^{-20}$ eV, induced decay of the vacuum gives by far the strongest constraint, as long as $\beta\kappa<0$. 

\section{Conclusion\label{sec:conclusion}}

In this article we explored the impact of technically natural-sized couplings on fifth force searches for new scalars.
Starting with the Yukawa coupling to matter that is responsible for the scalar fifth force, 
renormalization group effects generate all terms in the potential consistent with 
the symmetries. Self-interactions due to a non-trivial potential are known
to cause screening effects that reduce the effective fifth force. Aside from the standard screening behavior,
we showed that these terms can also cause enhancing effects. We also demonstrated how natural self-interactions lead to the Earth-LAGEOS constraint being unbounded above, thereby closing off the gaps that opened in other experimental coverage.

We showed that while screening effects due to quartic interactions limit the sensitivity of certain searches for long-range forces, they also enable new constraints. Because screening depends on the size and mass of the sources, the fundamentally EP-preserving fifth force becomes 
  effectively EP-violating. As an example, we showed how the recently discovered triple stellar system \triplesys can be used to constrain 
  this effective EP violation.

Technically natural-sized cubic interactions can lead to qualitatively different effects depending on their sign relative to fifth force matter coupling.
For one sign, the cubic terms give rise to screening, in analogy to the quartic case. This weakens certain constraints and introduces new ones due the effective EP violation.   
Large cubic interactions also modify long-distance behavior of these forces. 
Cubic terms change the vacuum structure of the scalar potential. As a result, the other sign of the cubic interaction (relative to the matter coupling) 
can lead to the enhancement of the fifth force charge of some sources. Moreover, stars have the capability of classically destabilizing our meta-stable minimum, resulting in some of the strongest bounds for long range fifth forces. Quantum tunneling also places a constraint, albeit a much weaker one.

Perhaps one of the most interesting aspects of a scalar fifth force lies in what a discovery would entail. We showed that the measurement of a new fifth force and the subsequent validation of its $1/r$ potential, would place bounds on higher-dimensional operators many orders of magnitude beyond the Planck scale. This would be the first real experimental test of quantum gravity. Because theorists have strong beliefs as to how quantum gravity corrections should behave, an experimental test of these properties would be either very reassuring or ground breaking.

\subsection*{Acknowledgements}
We would like to thank Asimina Arvanitaki, Prateek Agrawal, Asher Berlin, Zackaria Chacko, Patrick Draper, John Ellis, Kiel Howe, Junwu Huang, David McKeen, David Morrissey, Raman Sundrum and Natalia Toro for useful discussions. We would also like to thank Francisco Villatoro's \href{https://francis.naukas.com/2018/07/04/la-universalidad-de-la-caida-libre-en-relatividad-general-en-un-sistema-estelar-triple/}{talking mule} for alerting us to the recent stellar triple system constraint on EP-violating forces. N.B. and S.A.R.E. are supported by the U.S. Department of Energy under Contract No. DE-AC02-76SF00515. S.A.R.E. is also supported by the SNF through grant P2SKP2 171767.  A.H. is supported in part by the NSF under Grant No. PHY-1620074 and by the Maryland Center for Fundamental Physics (MCFP).

\bibliographystyle{JHEP}
\bibliography{biblio}
\end{document}